\documentclass[11pt,onecolumn,amssymb]{IEEEtran}
\usepackage[caption=false,font=footnotesize]{subfig}
\usepackage{graphicx}
\usepackage{booktabs}
\usepackage{amsmath}
\usepackage{amssymb}
\usepackage{midfloat}
\usepackage{tabularx}
\usepackage{bm}
\usepackage{color, cite}
\usepackage{algorithm}
\usepackage{array} 
\usepackage{mathtools}
\usepackage[noend]{algpseudocode}

\newtheorem{definition}{Definition}

\newtheorem{lemma}{Lemma}

\begin{document}

\title{Super-Resolution Delay-Doppler Estimation for OFDM Passive Radar}
\author{Le Zheng and Xiaodong Wang, \emph{Fellow}, \emph{IEEE},
\thanks{Le Zheng and Xiaodong Wang are in Electrical Engineering Department, Columbia University, New York, USA, 10027, e-mail: le.zheng.cn@gmail.com, wangx@ee.columbia.edu.}}
\maketitle
\begin{abstract}
In this paper, we consider the problem of joint delay-Doppler estimation of moving targets in a passive radar that makes use of orthogonal frequency-division multiplexing (OFDM) communication signals. A compressed sensing algorithm is proposed to achieve supper-resolution and better accuracy, using both the atomic norm and the $\ell_1$-norm. The atomic norm is used to manifest the signal sparsity in the continuous domain. Unlike previous works which assume the demodulation to be error free, we explicitly introduce the demodulation error signal whose sparsity is imposed by the $\ell_1$-norm. On this basis, the delays and Doppler frequencies are estimated by solving a semidefinite program (SDP) which is convex. We also develop an iterative method for solving this SDP via the alternating direction method of multipliers (ADMM) where each iteration involves closed-form computation. Simulation results are presented to illustrate the high performance of the proposed algorithm. 

\end{abstract}
\IEEEpeerreviewmaketitle

\section{Introduction}

Passive sensing has been employed in many applications, such as radar \cite{griffiths2005passive1,griffiths2005passive2}, underwater acoustics \cite{xerri2002passive,yocom2011bayesian}, wireless sensor networks \cite{zhang2015asymptotically} and seismology \cite{artman2006imaging}. As discussed in \cite{zhang2016joint}, passive radar systems, which detect and track targets of interest by utilizing illuminators of a communication system, have many benefits compared with the active radar systems. In particular, since the transmitter is not required, a passive radar is smaller and less expensive. Moreover, as shown in \cite{howland2005fm}, a passive radar uses ambient communication signals and has the capacity to operate without causing interference to existing wireless communication systems.

Challenges in implementing a passive radar system are mostly due to the usage of broadcast signals, which are not under control for illumination \cite{berger2010signal}. As the transmitted signals are not known a priori, a regular matched filter based receiver cannot be implemented easily. Moreover, the direct signal and clutters cause interference to the radar signal processing, making it hard for target detection and parameter estimation.  

Some passive radar systems do not demodulate the received signal. To eliminate the unwanted echoes in the surveillance channels (SCs), an additional separate channel, referred to as reference channel (RC), is introduced to collect the transmitted signal as a reference for passive detection \cite{liu2014two,tao2010direct}. The signal in the RC is regarded as a noisy template to implement an approximate matched filter. However, the RC is not very clean in many practical cases, and the performance of radar is significantly degraded when there are lots of interference, clutters and noise \cite{liu2014two}. 

To improve the performance of passive radar, one can make use of the structural information of the underlying communication signal. In particular, since the type of modulation is typically known, one can first estimate the data symbols by demodulation so that the matched filter can then be implemented \cite{colone2009multistage,berger2010signal}. As the demodulation provides better accuracy than directly using the signal in the RC, the detection and estimation performance of such radar systems can be improved. Moreover, as the data can be demodulated based on the received signal and the estimated channel states, the RC is not necessary and the complexity of the passive radar can be reduced.

Many modern wireless communications employ the orthogonal frequency-division multiplexing (OFDM) technique, that transmits data symbols over orthogonal subcarriers \cite{li2006orthogonal,bolcskei2006mimo,sen2011adaptive}. OFDM signals can be generated efficiently by using the fast Fourier transform (FFT), and have been used in passive radar applications, e.g., in \cite{zhao2012multipath}. Therefore, passive radar based on OFDM signals has received significant interest \cite{ketpan2015target,berger2010signal}. The illuminators that employ the OFDM signals includes digital video broadcasting (DVB) \cite{palmer2013dvb}, digital audio broadcasting (DAB) \cite{berger2010signal}, wireless local area network (LAN) \cite{falcone2010experimental,rzewuski2011system} and 4G wireless cellular \cite{salah2014experimental}. 

When considering signals modulated using OFDM, the broadcast signal can be treated as a special form of pulse-Doppler radar signal \cite{crouse2012time}. For some digital broadcast systems such as DVB/DAB, the communication can be considered as error free \cite{berger2010signal,choi2014approaches}. The targets can be searched using a bank of correlators tuned to the waveform given a certain Doppler shift and delay, i.e., a matched-filter bank. For better target resolution and clutter removal, the MUSIC and the compressed sensing (CS) based receivers have also been developed with higher computational complexity \cite{berger2010signal}.

Although several algorithms have been developed for passive radar with OFDM waveforms, the assumption of perfect demodulation does not hold in many wireless communication applications, which may lead to performance degradation. The demodulation error leads to an equivalent impulsive noise in the signal model, thereby increasing sidelobes of the clutters and strong targets. As the reflection of targets is usually small, it will cause strong interference for target detection and estimation.

An effective approach to passive radar signal processing is to exploit the sparsity properties of the signal. For example, the signal reflected by the target and clutters is a linear combination of several sinusoids. As the number of targets and clutters is small, the signal is sparse in the delay-Doppler domain. Moreover, under normal operating conditions, the demodulation error rate of a communication system is typically low. Therefore the demodulation error signal is also sparse.

With the development of sparse signal representation and later the compressed sensing (CS) theory \cite{donoho2006compressed}, which studies the recovery of a sparse signal from a number of linear measurements much less than its ambient dimension,
sparse methods have been developed for frequency recovery \cite{stankovic2013compressive}, computer vision \cite{zhang2016transfer,zhang2015cognitive} and denoising \cite{jokanovic2015reduced,studer2012recovery}. However, CS algorithms focuse on signals that can be sparsely represented by a finite discrete dictionary \cite{yang2014exact}, and therefore the frequencies of interest are assumed to lie on some fixed frequency grids. However, as the targets and clutters are usually specified by parameters in a continuous domain (delays and Doppler frequencies), the discretization usually leads to model mismatch and degradation in recovery \cite{chi2011sensitivity}.

To overcome the grid mismatch by the traditional sparsity-based methods, we apply the recently developed mathematical theory of continuous sparse recovery for super-resolution \cite{candes2013super,candes2014towards,tang2013compressed}. In \cite{candes2013super,candes2014towards}, the authors treat the complete data case and show that the frequencies can be exactly recovered via convex optimization once the separations between the frequencies are larger than certain threshold. In \cite{tang2013compressed,bhaskar2013atomic}, the result is extended to the problem of continuous frequency recovery from incomplete data based on the atomic norm minimization. Super-resolution based on atomic norm has many applications including direction of arrival estimation \cite{tan2014direction}, channel estimation \cite{chi2015compressive} and line spectral estimation \cite{tang2014robust}.

In this paper, we propose a new super-resolution algorithm for OFDM passive radar. We consider the case when the signal is sparse in the Fourier-domain. To estimate the signal from the corrupted measurements, we use the atomic norm constraint to formulate the sparsity of the signal in the delay-Doppler plane and use the $\ell_1$-norm to formulate the sparsity of the demodulation error signal. The parameter of the targets and clutters can be estimated by solving a semidefinite program (SDP) which is convex. To obtain a computationally efficient solution, we also develop an iterative algorithm based on the alternating direction method of multipliers (ADMM) \cite{boyd2011distributed} where each iteration involves closed-form computation. Extensive simulation results are provided to illustrate the performance advantage of the proposed algorithm over the existing super-resolution methods. 

The remainder of the paper is organized as follows. In Section II, we present the signal model of the passive radar. In Section III, we develop the proposed CS-based super-resolution receiver using both the atomic norm and the $\ell_1$-norm. Simulation results are presented in Section IV. Section V concludes the paper.

\section{System Descriptions}
\subsection{Signal Model}
We consider a passive radar system that exploits the OFDM communication signals, which is a multicarrier modulation scheme. The transmitted data is divided into multiple blocks. Suppose there are $N$ orthogonal subcarriers in each block and the frequency spacing of adjacent subcarriers is set as $\Delta f =1/T$. The duaration of each transmission block is $\bar T = T+T_\text{cp}$ where $T_\text{cp}$ is the length of the cyclic prefix. The transmitted signal in the $m$-th data block is given by
\begin{eqnarray}
\label{eq:xm}
x_m(t) = \sum_{n=0}^{N-1} s_m(n) e^{i 2\pi n \Delta f t } \xi(t - m \bar T), ~~ m \bar T - T_\text{cp} \leq t \leq m \bar T + T,
\end{eqnarray}
where $s_m(n)$, $n=0,1,...,N-1$, are data symbols in the $m$-th block; and
\begin{eqnarray}
\xi(t) = \left\{ \begin{array}{l}
1, ~ ~ t \in [-T_\text{cp},T],\\
0, ~ ~ \text{otherwise}.
\end{array} \right.
\end{eqnarray}
The transmitted signal over $M$ blocks can be expressed as $x(t)=\sum_{m=0}^{M-1} x_m(t)$, $- T_\text{cp} \leq t \leq (M-1) \bar T + T$. The baseband signal is then upconverted for transmission, i.e., $\tilde x(t)=e^{i 2 \pi f_c t}x(t)$ where $f_c$ denotes the carrier frequency.

On the receiver side, we assume that down-conversion has been performed and only consider the baseband signal in the following. We model the reflection of targets, clutters and the direct path as multipaths characterized by delays and Doppler frequencies. Suppose there are $K$ paths in the surveillance area. The delay and Doppler frequency of the $k$-th path are denoted as $\tau_k$ and $f_k$, respectively. Its attenuations including path loss, reflection and processing gains are characterized by a complex coefficient $A_k$. The received signal is then given by
\begin{eqnarray}
\label{eq:y}
y(t) &=& \sum_{k=1}^K A_k e^{i2\pi f_k t} x(t-\tau_k) + w(t), \nonumber \\
&=& \sum_{m=0}^{M-1} \sum_{k=1}^K A_k e^{i2\pi f_k t} x_m(t-\tau_k) + w(t), \nonumber \\
&\approx& \sum_{m=0}^{M-1} \sum_{k=1}^K A_k e^{i2\pi f_k m \bar T} x_m(t-\tau_k) + w(t),
\end{eqnarray}
where $w(t)$ is the additive noise; the approximation in the last line follows from \cite{berger2010signal}: since $f_k \bar T \ll 1$, the phase rotation within one OFDM block can be approximated as constant. 

A rectangular window of length $T$ is used at the receiver. Taking the Fourier transform on the received signal in the $m$-th block, the signal in the $n$-th subcarrier can be formulated as
\begin{eqnarray}
\label{eq:rk}
r_m(n) = \int_{m \bar T}^{m \bar T + T} e^{-i 2\pi n \Delta f t} y(t)dt.
\end{eqnarray}
Plugging \eqref{eq:xm} and \eqref{eq:y} into \eqref{eq:rk}, we have
\begin{eqnarray}
\label{eq:rk2}
r_m(n) &=& \int_{m \bar T}^{m \bar T + T} e^{-i 2\pi n \Delta f t} \sum_{k=1}^{K} A_k e^{i 2\pi m f_k \bar T} x_m(t-\tau_k) dt + \int_{m \bar T}^{m \bar T + T} e^{-i 2\pi n \Delta f t} w(t)dt \nonumber \\
&=& \int_{m \bar T}^{m \bar T + T} e^{-i 2\pi n \Delta f t} \sum_{k=1}^{K} A_k e^{i 2\pi m f_k \bar T} \sum_{l=0}^{N-1} s_m(l) e^{i 2\pi l \Delta f (t - \tau_k) } dt + \int_{m \bar T}^{m \bar T + T} e^{-i 2\pi n \Delta f t} w(t)dt \nonumber \\
&=& s_m(n) \sum_{k=1}^{K} A_k e^{i 2\pi m f_k \bar T} \int_{m \bar T}^{m \bar T + T} e^{- i 2\pi n \Delta f  \tau_k } dt + \int_{m \bar T}^{m \bar T + T} e^{-i 2\pi n \Delta f t} w(t)dt,
\end{eqnarray}
where $\xi(t)$ is dropped in the second line because it equals to 1 in the integration interval; the terms for $l \ne n$ are dropped in the third line because $\int_{m \bar T}^{m \bar T + T} e^{-i 2 \pi (n-l) \Delta f t} dt = 0$ for $l \ne n$. For notational simplicity, we define $\alpha_k = A_k T $, $\phi_k \triangleq f_k \bar T \in [0,1)$ and $\psi_k \triangleq \Delta f \tau_k \in [0,1)$, then $r_m(n)$ can be rewritten as
\begin{eqnarray}
\label{eq:rk3}
r_m(n) = \sum_{k=1}^{K} \alpha_k e^{i(2\pi m \phi_k - 2\pi n \psi_k)} s_m(n)  + v_m(n),
\end{eqnarray}
where 
\begin{eqnarray}
v_m (n) \triangleq \int_{m \bar T}^{m \bar T + T} e^{-i 2\pi n \Delta f t} w(t) dt.
\end{eqnarray}
It is assumed that $v_m (n)$ is a complex Gaussian variable with zero mean and variance $\sigma^2$, i.e., $v_m (n) \sim {\cal CN}(0,\sigma^2)$. In practice, the transmitted symbol $s_m(n)$ is unknown, but can be estimated by demodulation. For example, a passive radar often utilizes a reference channel at the receiver to collect a direct-path (transmitter-to-receiver) signal and estimate the symbols \cite{liu2014two}. For the convenience of our analysis, we define
\begin{eqnarray}
\label{eq:z}
z_m(n) &=& \sum_{k=1}^{K} \alpha_k e^{i(2\pi m \phi_k - 2\pi n \psi_k)}, \\
\label{eq:e}
e_m(n) &=& r_m(n) - \hat s_m(n) z_m(n) - v_m(n),
\end{eqnarray}
where $\hat s_m(n)$ is the estimated symbol. Obviously, $e_m(n) = 0$ if $\hat s_m(n) = s_m(n)$, and $e_m(n) \ne 0$ if $\hat s_m(n) \ne s_m(n)$. In this paper, we assume that the passive radar system only performs demodulation of the data symbols, but not the forward error correction (FEC) decoding \cite{ji2005rate}. Although the FEC can reduce the bit error rate, it also increases the complexity of the radar signal processing. In particular, powerful FEC codes, such as Turbo codes or LDPC codes, require the encoding of a large data block size, which means a large delay in passive radar systems and is not preferable. Moreover, for security or privacy reasons, the code book may not be available to the passive radar system, so the decoding may not be possible. Note that the demodulation error rate of a communication system is typically low under normal operating conditions. Therefore, most of the $e_m(n)$ equal to 0 and the noise caused by the demodulation error is sparse.

Denote the steering vectors $\bm b(\phi) = [1, e^{i 2 \pi \phi},...,e^{i 2 \pi (M-1) \phi}]^T$ and $\bm g(\psi) = [1, e^{i 2 \pi \psi},...,e^{i 2 \pi (N-1) \psi}]^T$. Correspondingly, the response matrices are defined as $ \bm B(\bm \phi)= [\bm b(\phi_1), \bm b(\phi_2),..., \bm b(\phi_K)]$ and $\bm G (\bm \psi) = [\bm g(\psi_1),\bm g(\psi_2),...,\bm g(\psi_K)]$ where $\bm \phi = [\phi_1, \phi_2,...,\phi_K]^T$ and $\bm \psi = [\psi_1, \psi_2,...,\psi_K]^T$. Then \eqref{eq:z} and \eqref{eq:e} can be written in a matrix form
\begin{eqnarray}
\label{eq:R}
\bm R &=& \bm{\hat S} \odot \bm{Z} + \bm E + \bm V \nonumber \\
&=& \bm{\hat S} \odot \left( {\bm B}(\bm \phi) \text{diag}(\bm{\alpha}) \bm{G}(\bm \psi)^H \right) + \bm E + \bm V,
\end{eqnarray}
where $\bm{\alpha} = [\alpha_1,\alpha_2,..., \alpha_K]^T$; $\odot$ denotes the Hadamard product; $\bm{\hat S} \in \mathbb{C}^{M \times N}$, $\bm R \in \mathbb{C}^{M \times N}$, $\bm{Z} \in \mathbb{C}^{M \times N}$, $\bm E \in \mathbb{C}^{M \times N}$ and $\bm V \in \mathbb{C}^{M \times N}$ are matrices whose $(m,n)$-th element are $\hat s_m(n)$, $r_m(n)$, $z_m(n)$, $e_m(n)$ and $v_m(n)$, respectively. To explore the structure of the signal, we vectorize $\bm R$ and obtain
\begin{eqnarray}
\label{eq:barr}
\bm{\bar r} &=& \bm{\tilde S}\bm{\bar z} + \bm{\bar e} + \bm{\bar v} \nonumber \\
&=& \bm{\tilde S} \left(\bm G(\bm \psi)^* \circ \bm B(\bm \phi) \right) \bm{\alpha} + \bm{\bar e} + \bm{\bar v},
\end{eqnarray}
where $\bm{\tilde S} \triangleq \text{diag}(\text{vec}(\bm{\hat S})) \in \mathbb{C}^{MN \times MN}$, $\bm{\bar z} \triangleq \text{vec}(\bm{Z}) \in \mathbb{C}^{MN \times 1}$, $\bm{\bar e} \triangleq \text{vec}(\bm E) \in \mathbb{C}^{MN \times 1}$ and $\bm{\bar v} \triangleq \text{vec}(\bm V) \in \mathbb{C}^{MN \times 1}$; $\circ$ is the Khatri-Rao product; $\left(\bm G(\bm \psi)^* \circ \bm B(\bm \phi) \right) \in \mathbb{C}^{MN \times K}$ is a matrix whose $k$-th column has the form $\bm g(\psi_k)^* \otimes \bm b(\phi_k)$ where $\otimes$ is the Kronecker product.

The problem is to jointly estimate the unknown parameters $\bm \alpha$, $\bm \phi$ and $\bm \psi$ from the noisy observations $\bm{\bar r}$. With these estimates, we can then compute the reflections $\alpha_k$, delays $\tau_k$ and Doppler frequencies $f_k$ of the targets in the surveillance area.

\subsection{Existing Methods}

Several algorithms have been developed for the case of perfect demodulation. In \cite{berger2010signal}, a method based on matched-filtering and conventional FFT processing is given first, that suffers from low resolution. For better target resolution and clutter removal, super-resolution methods based on MUSIC and CS are also proposed in \cite{berger2010signal}, which are briefly described next. It is worth noting that \cite{berger2010signal} assumes that the data symbols $s_m(n)$ are with unit magnitude, i.e., phase-shift keying (PSK) modulation. However, typically in OFDM communications the quadrature amplitude modulation (QAM) is employed, where the symbols have different magnitudes. In the following, we discuss the two super-resolution methods for the general modulation.

\subsubsection{Super-resolution Receiver based on MUSIC}
When $\bm{\bar e} = \bm 0$, i.e., $s_m(n) = \hat s_m(n)$, the delays and Doppler shifts of the paths can be directly estimated. As several snapshots of the waveform are required for MUSIC, the spatial smoothing should be applied first. Specifically, the steering vector for the MUSIC algorithm is defined as
\begin{eqnarray}
\bm a'(\phi,\psi) = \text{vec}(\bm A'(\phi,\psi)) \in \mathbb{C}^{M'N'\times 1},
\end{eqnarray}
where
\begin{eqnarray}
\bm A'(\phi,\psi) = \left[ {\begin{array}{*{20}{c}}
	1&{{e^{ - i2\pi \psi }}}& \cdots &{{e^{ - i2\pi (N' - 1)\psi }}}\\
	{{e^{i2\pi \phi }}}&{{e^{i2\pi (\phi  - \psi )}}}& \cdots &{{e^{i2\pi [\phi  - (N' - 1)\psi ]}}}\\
	\vdots & \vdots & \ddots & \vdots \\
	{{e^{i2\pi (M' - 1)\phi }}}&{{e^{i2\pi [(M' - 1)\phi  - \psi ]}}}& \cdots &{{e^{i2\pi [(M' - 1)\phi  - (N' - 1)\psi ]}}}
	\end{array}} \right] \in \mathbb{C}^{M' \times N'}
\end{eqnarray}
with $N'<N$ and $M'<M$. Suppose there are $N_\text{snap} = (N-N'+1)(M-M'+1)$ snapshots after smoothing. Grouping the received signal and stacking the columns of each matrix on top of each other, we can build an observation matrix
\begin{eqnarray}
\bm Y = \left[\text{vec}(\bm R'_{0,0}), ...,\text{vec}(\bm R'_{N-N',0}),\text{vec}(\bm R'_{0,1}),..., \text{vec}(\bm R'_{N-N',M-M'}) \right] \in \mathbb{C}^{M'N' \times N_\text{snap}},
\end{eqnarray}
where
\begin{eqnarray}
\bm R_{m,n}' = \left[ {\begin{array}{*{20}{c}}
	R_m(n)&{R_m(n+1)}& \cdots &{R_m(n+N'-1)}\\
	{R_{m+1}(n)}&{R_{m+1}(n+1)}& \cdots &{R_{m+1}(n+N'-1)}\\
	\vdots & \vdots & \ddots & \vdots \\
	{R_{m+M'-1}(n)}&{R_{m+M'-1}(n+1)}& \cdots &{R_{m+M'-1}(n+N'-1)}
	\end{array}} \right] \in \mathbb{C}^{M' \times N'},
\end{eqnarray}
with $R_{m}(n) = \frac{r_{m}(n)}{\hat s_m(n)}$ for $m=0,1,...,M-M'$ and $n=0,1,...,N-N'$.

With the observation matrix after spatial smoothing, the signal and noise subspaces can be obtained via an SVD $\bm Y = \bm F \bm D \bm H$ where $\bm F \in \mathbb{C}^{M'N' \times M'N'}$ and $\bm H \in \mathbb{C}^{N_\text{snap} \times N_\text{snap}}$. Suppose $\bm F = [\bm F^{(s)}, \bm F^{(n)}]$ in which $\bm F^{(s)}$ corresponds to the signal subspace and $\bm F^{(n)}$ corresponds to the noise subspace. In practice, the signal subspace can be determined by finding the eigenvectors corresponding to the largest eigenvalues. The MUSIC spectrum can be given in terms of the noise subspace 
\begin{eqnarray}
f_\text{MUSIC} (\phi,\psi) = \frac{1}{\bm a'(\phi,\psi)^H (\bm F^{(n)})^H \bm F^{(n)} \bm a'(\phi,\psi)}.
\end{eqnarray}
The two-dimensional MUSIC (2D-MUSIC) method can estimate the targets and clutters by locating the poles in the spectrum. The amplitudes of the targets and clutters can then be estimated by the least-squares (LS) method given the delays and Doppler frequencies. 

The MUSIC algorithm has been popular for its good resolution and accuracy in frequency estimation \cite{naha2015determining}. However, it is also reported that the optimization-based method leveraging sparsity-inducing norms can outperform the MUSIC algorithm in noisy environments \cite{bhaskar2013atomic,yang2014exact}. Moreover, the MUSIC algorithm is designed to allow for small Gaussian-like perturbations to the data and hence their performance degrades gracefully when such noise is present. In passive radar, due to the demodulation error, some impulsive noise are present in the data, which can degrade the performance severely.

\subsubsection{Super-resolution Receiver based on Compressed Sensing}

In addition to the subspace algorithms such as MUSIC, the parameters can also be estimated by the CS algorithm. Obviously, jointly estimating $\bm \phi$, $\bm \psi$ and $\bm \alpha$ in \eqref{eq:barr} is a non-linear problem. It can be linerized by using an overcomplete dictionary matrix $ \bm C' = (\bm{G'}^* \circ \bm{B'})$ where
\begin{eqnarray}
\bm B'= [\bm b(\phi_1'), \bm b(\phi_2'),..., \bm b(\phi_{\tilde M}')] \in \mathbb{C}^{M \times \tilde M}, \\
\bm G' = [\bm g(\psi_1'),\bm g(\psi_2'),...,\bm g(\psi_{\tilde N}')]  \in \mathbb{C}^{N \times \tilde N},
\end{eqnarray}
with $\{ \phi_m' \}$ and $\{ \psi_n' \}$ denoting sets of uniformly spaced frequency points. Define $\tilde L = \tilde M \tilde N$ as the number of columns of $\bm C'$ where $\tilde M \geq M$ and $\tilde N \geq N$. For sufficiently large $\tilde M$ and $\tilde N$, the frequency spectrum is densely sampled. Let $\bm \alpha' \in \mathbb{C}^{\tilde L \times 1}$ be the sparse vector whose non-zero elements correspond to $\bm \alpha$ in \eqref{eq:barr}. Then the non-linear joint estimation problem reduces to a linear parameter estimation problem, i.e., to estimate the linear amplitude vector $\bm \alpha'$ under a sparsity constraint:
\begin{eqnarray}
\label{eq:cs}
\bm{\hat \alpha'} = \arg\min_{\bm \alpha} \frac{1}{2}\| \bm{\bar r} - \bm{\tilde S} \bm{C}' \bm{\alpha}' \|_2^2 + \gamma\|\bm \alpha'\|_1
\end{eqnarray}
where $\| \cdot \|_1$ denotes the $\ell_1$-norm, $\gamma$ is a weighting parameter determining the sparsity of the reconstruction. As \eqref{eq:cs} is convex, the amplitudes can be estimated efficiently with convex optimization algorithms \cite{boyd2004convex}. The delays and Doppler frequencies can be identified by locating the non-zero entries of $\bm{\hat \alpha'}$.

The CS algorithm based on $\ell_1$-minimization (CS-L1) is capable of super-resolving the spectrum of the multi-sinusoidal signal under certain conditions of the sensing matrix $\bm C'$ \cite{donoho2006compressed}. However, the spectrum of interest is discretized into a number of grids, and the targets may not exactly reside on the grids. The off-grid targets can lead to mismatches in the model and deteriorate the performance significantly \cite{tan2014direction,chi2011sensitivity,tan2014joint}. Furthermore, the CS receiver in \eqref{eq:cs} does not consider the demodulation error, which can lead to further performance degradation.

\section{CS-based High-resolution Passive Radar using Atomic Norm}
\subsection{Atomic Norm Formulation}
As has been explained in the previous section, the demodulation errors lead to the impulsive noise in the measurements, and will degrade the performance of the super-resolution receiver based on MUSIC or CS. From \eqref{eq:barr}, $\bm{\bar r}$ is a function of scattering coefficients $\alpha_k$, delays $\tau_k$ and Doppler frequencies $f_k$. To reduce the impact of the impulsive noise, we will need to exploit the following two types of sparsity in the problem:
\begin{enumerate}
	\item The reflection of targets and clutters is a combination of complex sinusoids of different frequencies, and the number of frequencies is much smaller than the number of measurements, i.e., $K \ll MN$.
	\item The number of mistakenly demodulated symbols is much smaller than the number of measurements.
\end{enumerate}
Denote the number of mistakenly demodulated symbols as $J$, then we have $\|\bm{\bar e}\|_0 = J$. Although the $\ell_0$-norm constraint enhances the sparsity of the solution, it results in a non-convex optimization problem which is NP-hard. Therefore, we use the $\ell_1$-norm regularization instead, i.e., $\|\bm{\bar e}\|_1 = \sum_{l=1}^{MN} |\bar e(l)|$.

The signal $\bm{\bar z} $ is a linear combination of complex sinusoids with arbitrary phases, where the frequencies do not fall onto discrete grids. Therefore, it cannot be directly formulated by using the $\ell_1$-norm. To solve the off-grid problem, we use the atomic norm to build the sparse representation of the signal. Define an atom as $\bm a(\phi,\psi) = \bm g(\psi)^* \otimes \bm b(\phi)$, and the set of atoms is defined as the collection of all normalized 2D complex sinusoids: ${\cal A} = \{ \bm a(\phi,\psi) : \phi \in [0,1), \psi \in [0,1) \}$. Obviously, we have $\bm{\bar z} = \sum_{k=1}^K \alpha_k \bm a(\phi_k,\psi_k)$ from \eqref{eq:z}.
\begin{definition}
	The 2D atomic norm for $\bm{\bar z} \in \mathbb{C}^{MN \times 1}$ is 
	\begin{eqnarray}
	\| \bm{\bar z} \|_{\cal A} = \mathop {\inf }\limits_{\scriptstyle{\phi_k} \in [0,1), \scriptstyle{\psi_k} \in [0,1) \hfill\atop
		\scriptstyle{\alpha_k} \in \mathbb{C}\hfill} \left\{ {\left. {\sum\limits_k {|{\alpha_k}|} } \right| \bm{\bar z} = \sum\limits_k {{\alpha_k}{\bm a}({\phi_k,\psi_k})}} \right\}.
	\end{eqnarray}
\end{definition}

The atomic norm can enforce sparsity in the atom set $\cal A$ \cite{tang2013compressed}. On this basis, an optimization problem will be formulated for the estimation of the signal frequencies. For the convenience of calculation, we will use the following equivalent form of the atomic norm for the atom set $\cal A$ \cite{yang2016vandermonde}:
\begin{eqnarray}
\label{eq:atomic}
\| \bm{ \bar z} \|_{\cal A} = \mathop {\inf }\limits_{\bm U, t } \left\{ \begin{array}{l}
\frac{1}{2MN}{\rm{Tr}}({\cal T}(\bm U)) + \frac{t }{2},\\
{\rm s.t.} \left[ {\begin{array}{*{20}{c}}
	{{\cal T}(\bm U)}& \bm{\bar z}\\
	{{\bm{\bar z}^H}}& t 
	\end{array}} \right] \succeq 0
\end{array} \right\},
\end{eqnarray}
where $\rm{Tr}(\cdot)$ is the trace of the input matrix; $\bm U$ is a $(2M-1) \times (2N-1)$ matrix defined as
\begin{eqnarray}
\label{eq:U}
\bm U = [\bm u_{-N+1},\bm u_{-N+2},...,\bm u_{N-1}]
\end{eqnarray}
with 
\begin{eqnarray}
\label{eq:u}
\bm u_l=[u_l(-M+1),u_l(-M+2),...,u_l(M-1)]^T;
\end{eqnarray}
${\cal T}(\bm U)$ is a block Toeplitz matrix defined as
\begin{eqnarray}
\label{eq:calT}
{\cal T}(\bm U) = \left[ {\begin{array}{*{20}{c}}
	{\text{Toep}(\bm u_0)}&{\text{Toep}(\bm u_{-1})}&{\cdots}&{\text{Toep}(\bm u_{-N+1})}\\
	{\text{Toep}(\bm u_1)}&{\text{Toep}(\bm u_0)}&{\cdots}&{\text{Toep}(\bm u_{-N+2})}\\
	{\vdots}&{\vdots}&{\ddots}&{\vdots}\\
	{\text{Toep}(\bm u_{N-1})}&{\text{Toep}(\bm u_{N-2})}&{\cdots}&{\text{Toep}(\bm u_0)}
	\end{array}} \right],
\end{eqnarray}
where $\text{Toep}(\cdot)$ denotes the Toeplitz matrix whose first column is the last $M$ elements of the input vector. More specifically, we have
\begin{eqnarray}
\label{eq:T}
\text{Toep}(\bm u_l) = \left[ {\begin{array}{*{20}{c}}
	{u_{l}(0)}&{{u_{l}(-1)}}&{\cdots}&{{u_{l}(-M+1)}}\\
	{u_{l}(1)}&{u_{l}(0)}&{\cdots}&{u_{l}(-M+2)}\\
	{\vdots}&{\vdots}&{\ddots}&{\vdots}\\
	{u_{l}(M-1)}&{u_{l}(M-2)}&{\cdots}&{u_{l}(0)}
	\end{array}} \right]
\end{eqnarray}
for $l=-N+1,-N+2,...,N-1$.

Using the atomic norm, we can formulate the following optimization problem based on \eqref{eq:z}:
\begin{eqnarray}
\label{eq:atomicnorm}
(\bm{\hat z},\bm{\hat e}) = \mathop {\arg \min }\limits_{\bm{\bar z},\bm{\bar e}} \frac{1}{2}{\left\| {{\bm{\bar r}} - \bm{\bar e} - \bm{\tilde S}\bm{\bar z}} \right\|_2^{2}} + \lambda \| \bm{\bar z} \|_{\cal A} + \mu \| \bm{\bar e} \|_1, 
\end{eqnarray}
where $\lambda > 0$ and $\mu>0$ are the weight factors. In practice, we set $\lambda \backsimeq \sigma \sqrt{MN \log(MN)}$ and $\mu \backsimeq \sigma \sqrt{\log(MN)}$. Applying \eqref{eq:atomic}, then \eqref{eq:atomicnorm} can be transformed to the following semidefinite program (SDP):
\begin{eqnarray}
\label{eq:SDP}
(\bm{\hat z},\bm{\hat e}) &=& \mathop {\arg \min }\limits_{\bm{\bar z}, \bm{\bar e}, \bm U, t}  \frac{1}{2}{\left\| {{\bm{\bar r}} - \bm{\bar e} - \bm{\tilde S} \bm{\bar z}} \right\|_2^{2}} + \frac{\lambda}{2MN} {\text{Tr}}\left( {\cal T}(\bm U) \right) + \frac{\lambda t}{2} + \mu \|\bm{\bar e}\|_1 \\
\text{s.t.} && \left[ {\begin{array}{*{20}{c}}
	{{\cal T}(\bm U)}& \bm{\bar z} \\
	{{\bm{\bar z}^H}}& t
	\end{array}} \right] \succeq 0. \nonumber
\end{eqnarray}
The above problem is convex, and can be solved by using a convex solver. We denote the solutions to \eqref{eq:SDP} as $\bm{\hat z} = [\hat z_1,\hat z_2,...,\hat z_{MN}]^T$ and $\bm{\hat e}=[\hat e_1,\hat e_2,...,\hat e_{MN}]^T$. The demodulation error can be identified by locating the non-zero elements in $\bm{\hat e}$. We name the super-resolution receiver given by \eqref{eq:atomicnorm} with $\mu \ne 0$ as the CS algorithm based on atomic norm and $\ell_1$-norm (CS-ANL1). 

\subsection{Example}

Consider a special case when the communication is error free and therefore the penalty term $\mu \|\bm{\bar e}\|_1$ in \eqref{eq:atomicnorm} is unnecessary. We name the super-resolution algorithm given by \eqref{eq:atomicnorm} with $\mu=0$ as the CS algorithm based on atomic norm (CS-AN). As the atomic norm can enforce sparsity in the continuous domain, its performance does not suffer from the off-grid problem. 

\begin{figure}
	\centering
	\subfloat[][]{\includegraphics[width=3in]{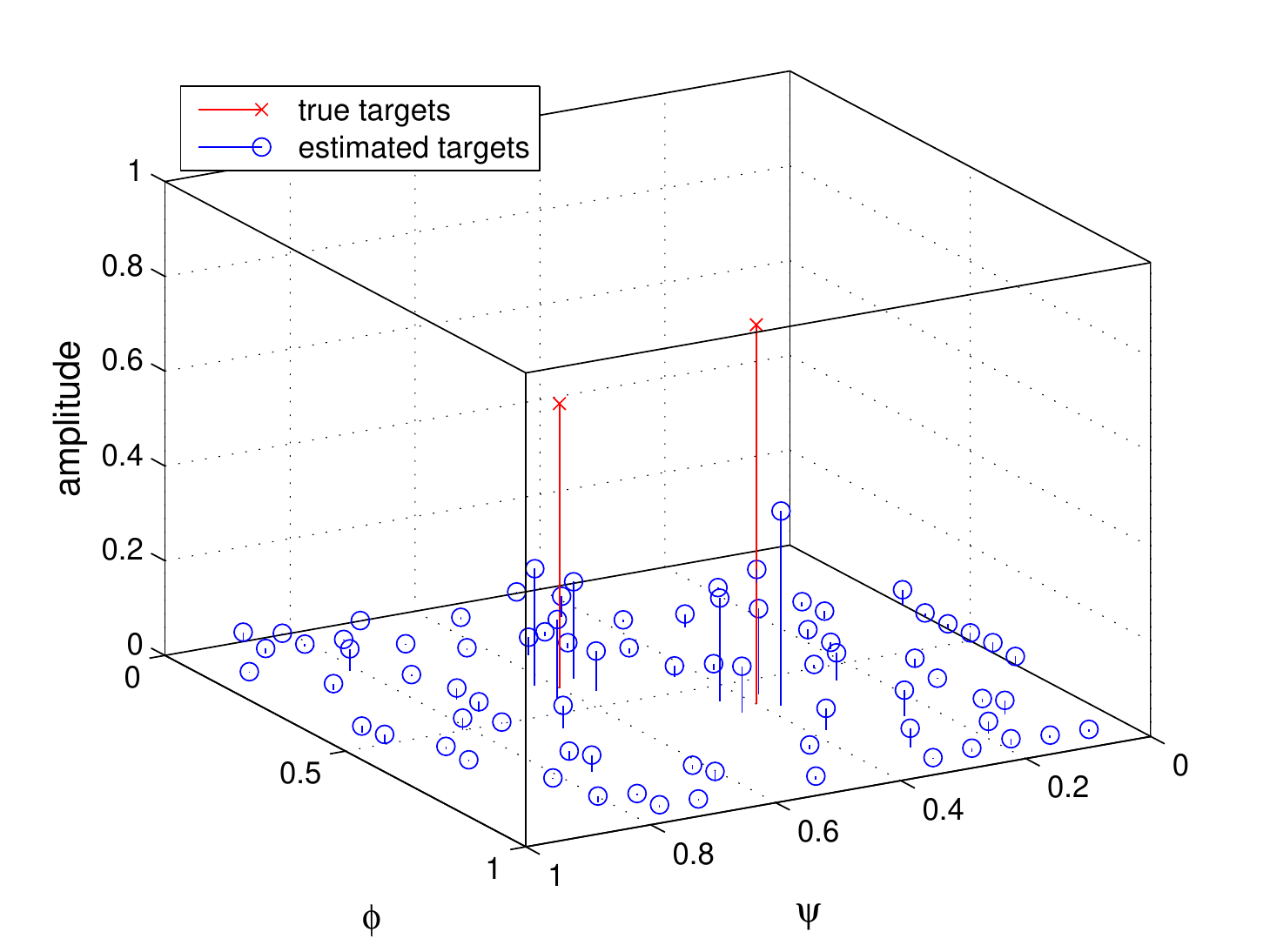}}
	\subfloat[][]{\includegraphics[width=3in]{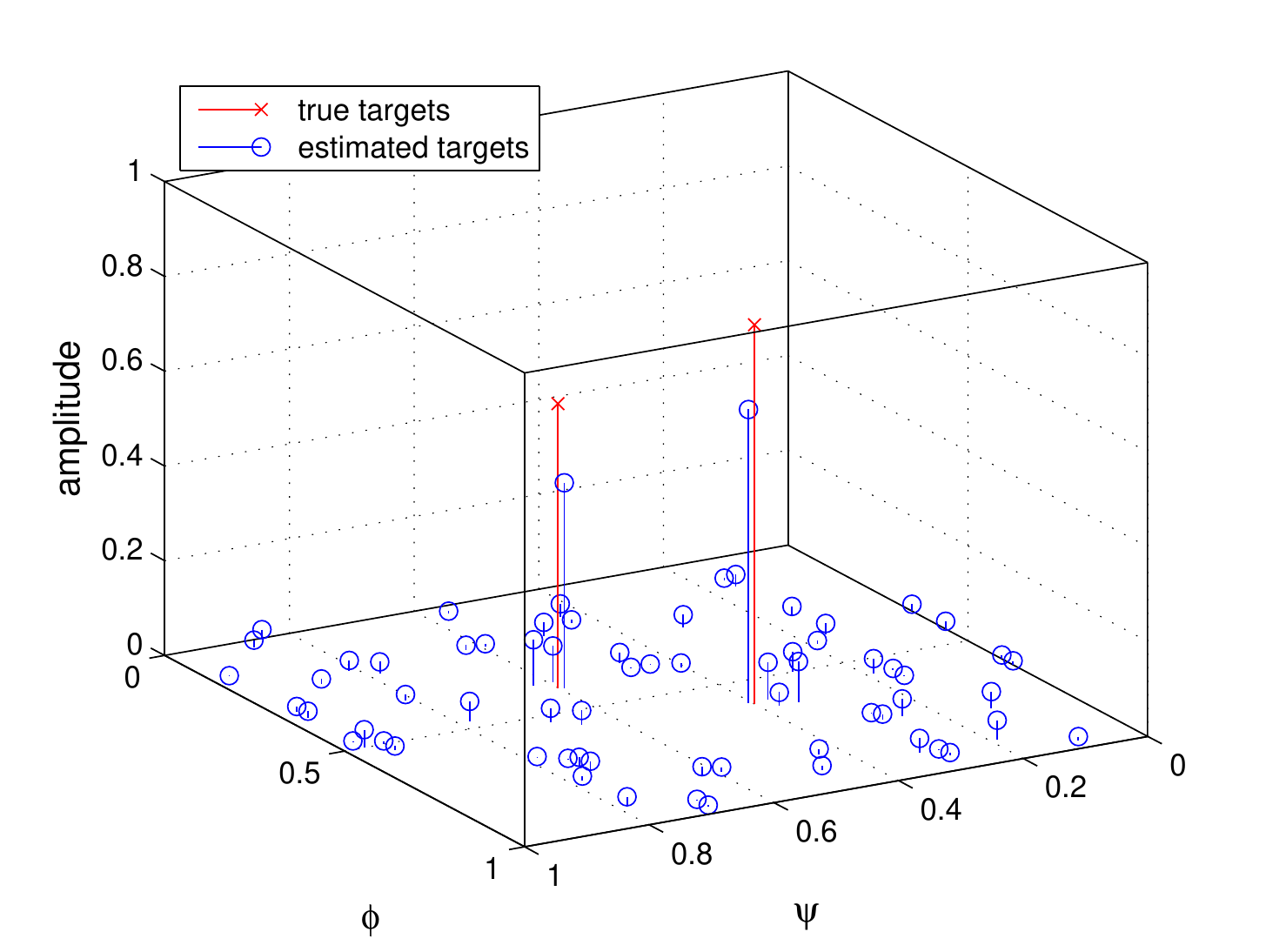}}
	
	\subfloat[][]{\includegraphics[width=3in]{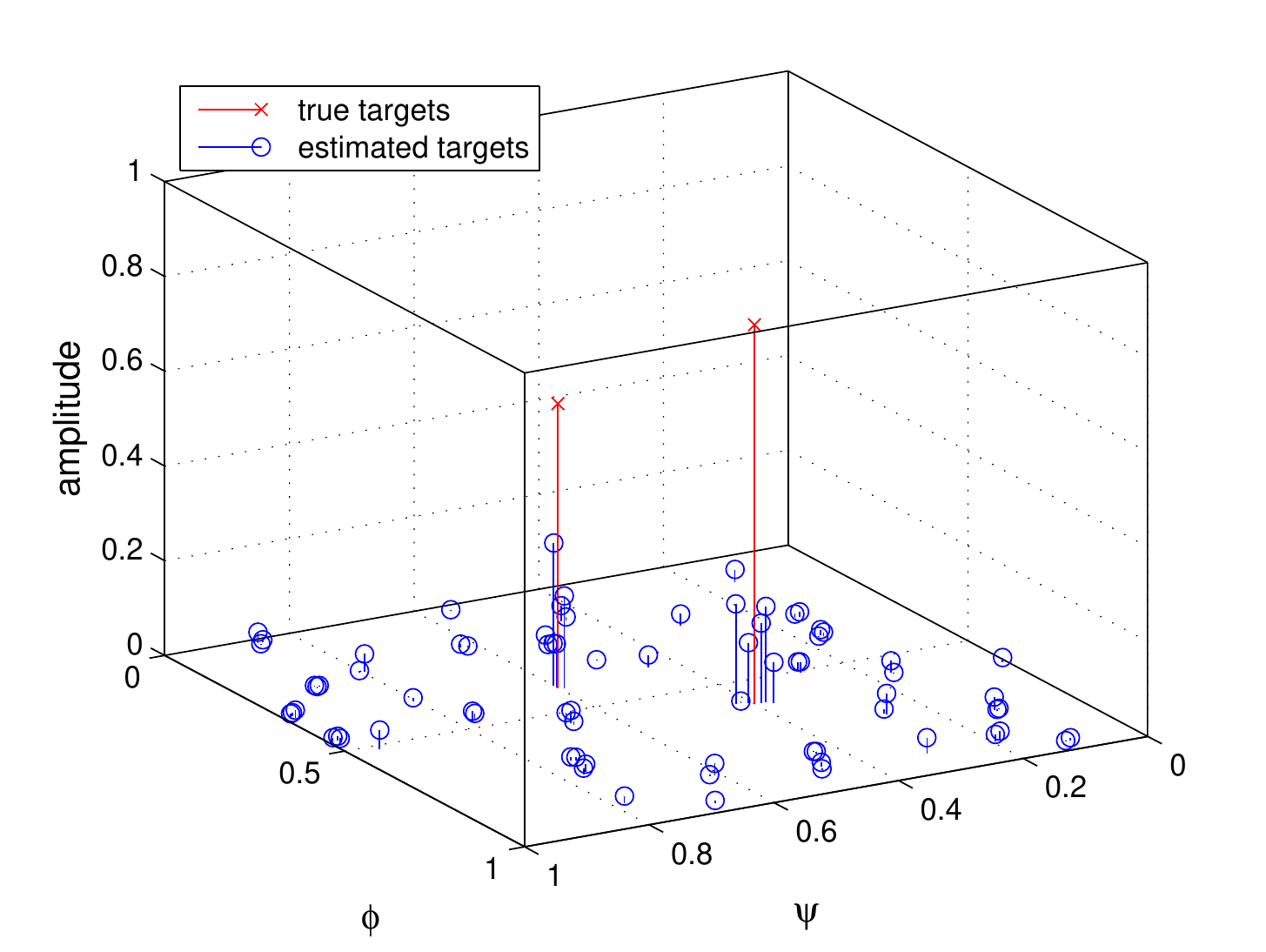}}
	\subfloat[][]{\includegraphics[width=3in]{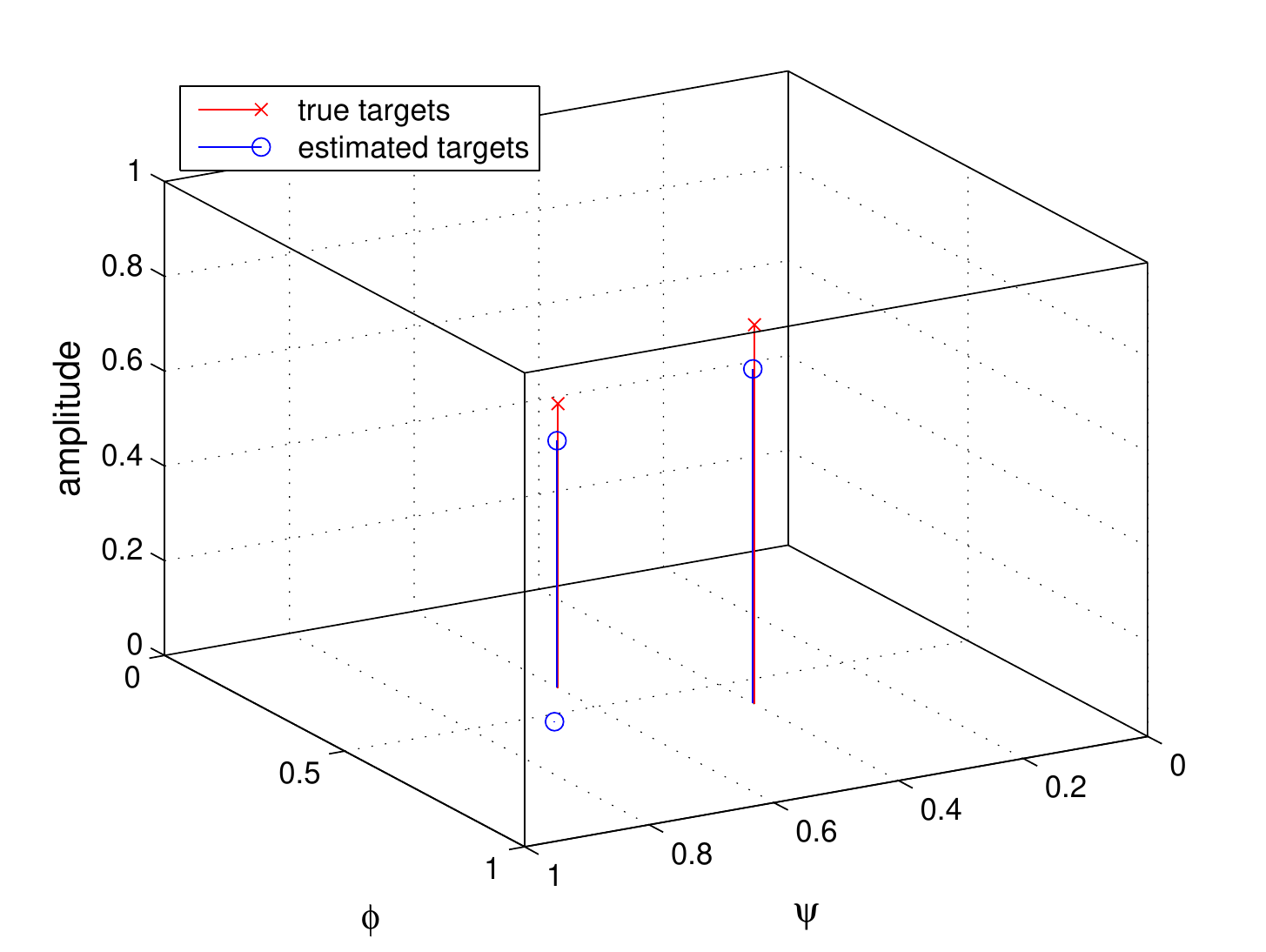}}
	
	\caption{Comparison between CS-AN algorithm and CS-L1 algorithm: (a) CS-L1 algorithm with $\kappa =2$, (b) CS-L1 algorithm with $\kappa =4$, (c) CS-L1 algorithm with $\kappa =16$ and (d) CS-AN algorithm.}
	\label{fig:example}
\end{figure}

In Fig. \ref{fig:example}, we compare the performance of the CS-AN algorithm and the CS-L1 algorithm with an example. The parameters are set as $M=N=8$ and $K = 2$. The variance of the noise is $\sigma^2 = 0.1$. The CS-AN receiver is implemented by solving \eqref{eq:atomicnorm} with $\lambda = \sigma \sqrt{MN \log (MN)}$ and $\mu = 0$. The delay and Doppler frequency estimation of the CS-AN receiver will be explained in Section III-C. The CS-L1 algorithm is implemented by solving \eqref{eq:cs} with CVX \cite{cvx}. The grid parameters of the CS algorithm are set as $\tilde M = \kappa M$ and $\tilde N = \kappa N$ with $\kappa = 2, 4, 16$. The weighting parameter of the CS-L1 algorithm is set as $\gamma = 2\sigma \sqrt{2 \log(\tilde M \tilde N)}$.

As can be seen from Fig. \ref{fig:example}, the CS-L1 algorithm returns many false alarms due to the basis mismatch. Since the magnitudes of some weak targets are also small, it may be hard to distinguish the false alarms from the targets. With the increase of the grid density, the mismatch caused by descritization can be reduced, but the coherence of the sensing matrix $\bm{C}'$ increases. As a result, the targets are still split into many scatterers. The CS-AN algorithm also returns one false alarm caused by the measurement noise. However, it is able to identify the target frequencies accurately and the number of false alarms is far less than that of CS-L1 algorithm.

\subsection{Delay and Doppler Frequency Estimation}

The purpose of the passive radar is mainly to detect and track the targets. However, solving \eqref{eq:SDP} does not directly provide the estimates of $\phi$ and $\psi$, so the delays and Doppler frequencies of the scatterers are still unavailable. One way to estimate the delays and Doppler frequencies is using the 2D-MUSIC with $\bm{\hat z}$ as the input. Another approach is to obtain the estimates from the dual solution of the problem. Following \cite{tang2013compressed}, the dual problem of \eqref{eq:atomicnorm} is given by
\begin{eqnarray}
	\label{eq:dual}
	\max_{\bm \nu} &&{\left\langle {{(\bm{\tilde S}^H)^{-1} \bm \nu},\bm{\bar r}} \right\rangle _{\mathbb R}} - \frac{1}{2}\left\| {(\bm{S}^H)^{-1} \bm \nu} \right\|_2^2, \\
	\text{s.t.} && \|\bm \nu \|_{\cal A}^* \leq \lambda \nonumber, \\
	&& \left\| (\bm{\tilde S}^H)^{-1} \bm \nu \right\|_\infty \leq \mu \nonumber,
\end{eqnarray}
where $\bm \nu \in \mathbb{C}^{MN \times 1}$ is the dual variable, $\langle \bm \nu , \bm{r} \rangle_{\mathbb R} = \text{Re}(\bm{r}^H \bm \nu)$, and $\| \bm \nu \|_{\cal A}^* = \sup_{\| \bm z \|_{\cal A} \leq 1} \langle \bm \nu , \bm z \rangle_{\mathbb R}$ is the dual norm. Let $\bm{\hat \nu}=[\hat \nu_1,\hat \nu_2,...,\hat \nu_{MN}]^T$ be the solution to \eqref{eq:dual}. Solving the dual problem is equivalent to solving the primal problem, and most solvers can directly return a dual optimal solution when solving the primal problem. The following lemma can be used to identify the frequencies from a dual solution.
\begin{lemma} \cite{fernandez2016demixing}
	Suppose $\bm{\hat z} = \sum \limits_{k = 1}^{{\hat K}} {{\hat \alpha_k} \bm a({\hat \phi_k, \hat \psi_k})}$ is the primal solution, then the dual polynomial $Q(\phi,\psi)=\langle \bm{\hat \nu}, \bm a(\phi,\psi) \rangle$ satisfies
	\begin{eqnarray}
		\label{eq:dualc}
		&& Q(\hat \phi_k,\hat \psi_k) = \lambda \frac{\hat \alpha_k}{|\hat \alpha_k|}, k=1,2,...,\hat K,\\
		&& \hat \nu_j = (\tilde S_{j,j}^*)^{-1} \mu \frac{\hat e_j}{|\hat e_j|}, \forall \hat e_j \ne 0, j=1,2,...,MN,
	\end{eqnarray}
	where $\hat K$ is the number of estimated frequencies, $\tilde S_{j,j}$ is the $j$-th diagnal element of the matrix $\bm{\tilde S}$.
\end{lemma}

According to \eqref{eq:dualc}, the delays and Doppler frequencies of the signal can be obtained by identifying points where the dual polynomial has modulus $\lambda$. Moreover, the dual solution provides another way to detect the mistaken demodulation: in places where mistaken demodulation occurs, the amplitude of the dual solution equals to $\mu$.

\begin{figure*}
	\centering
	\subfloat[][]{\includegraphics[width=3.2in]{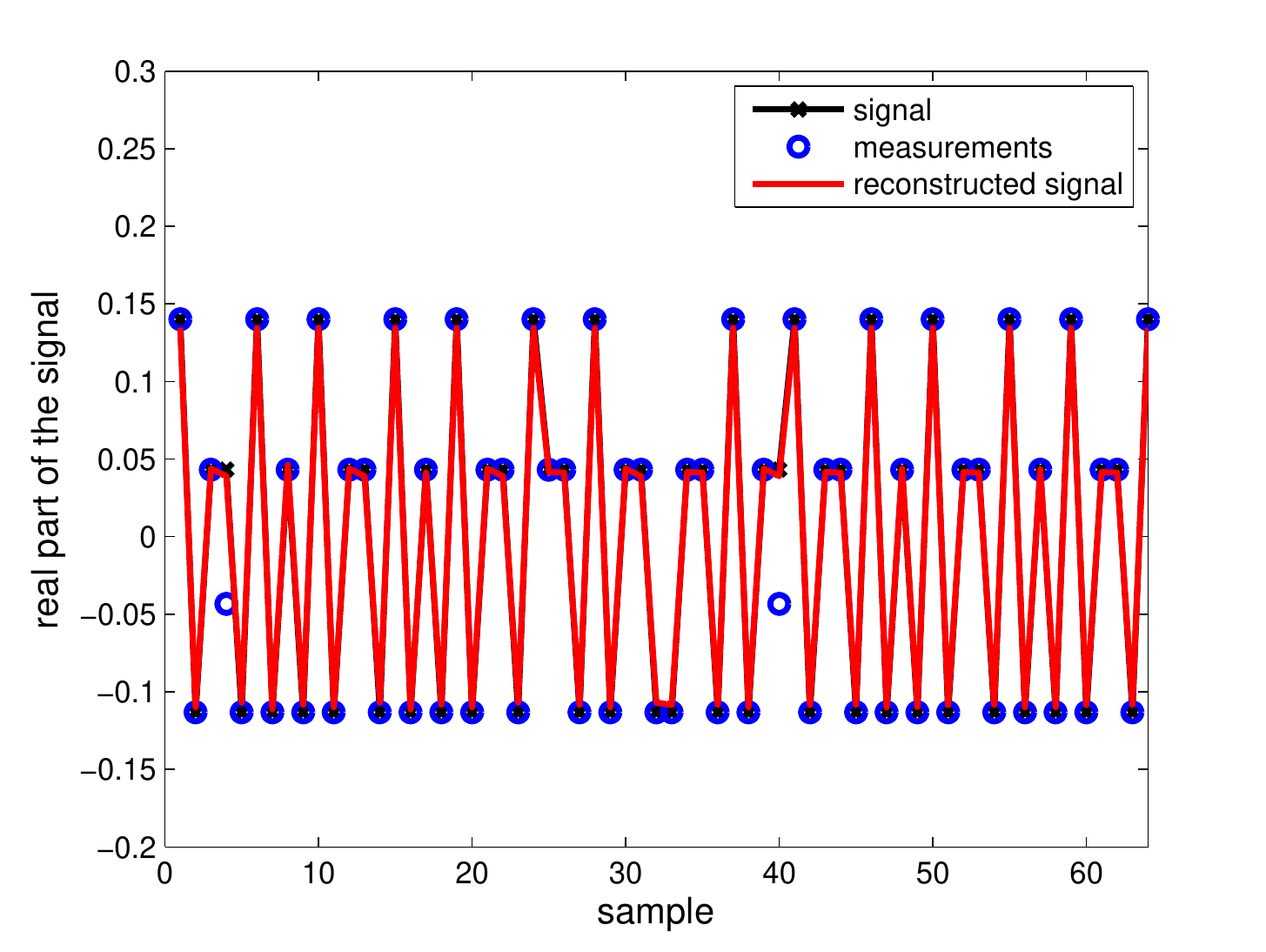}}
	\subfloat[][]{\includegraphics[width=3.2in]{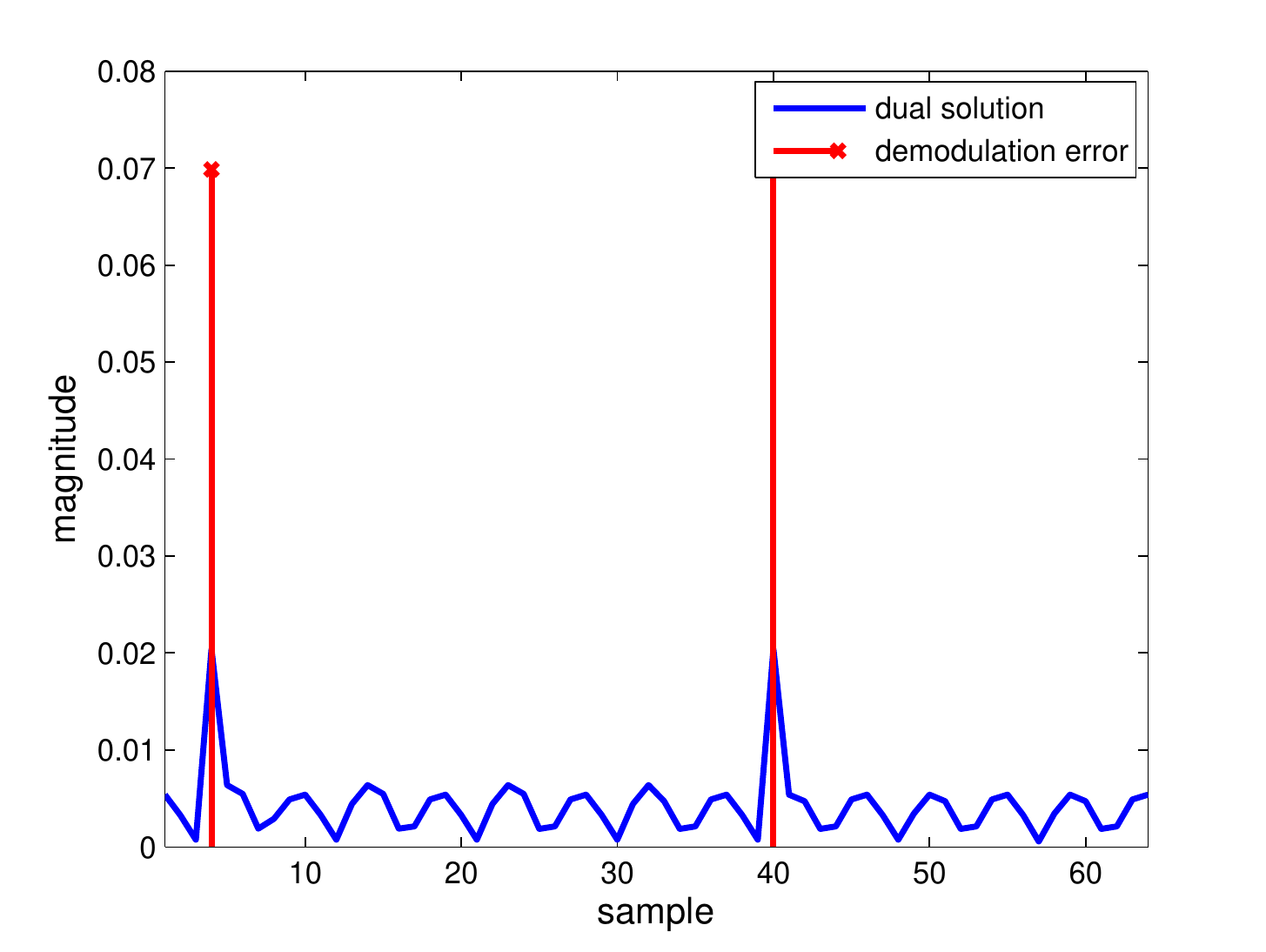}}
	
	\caption{(a) Plots of signal, measurements, and the estimated signal based on solving \eqref{eq:SDP}. (b) Plots of dual solution and the position of mistakenly demodulated symbols: the x-axis is the element index of $\bm {\hat \nu}$ and $\bm{\hat e}$, the y-axis is the magnitude of $\bm {\hat \nu}$ and $\bm{\hat e}$.}
	\label{fig:example2}
\end{figure*}

\begin{figure*}
	\centering
	\includegraphics[width=3.2in]{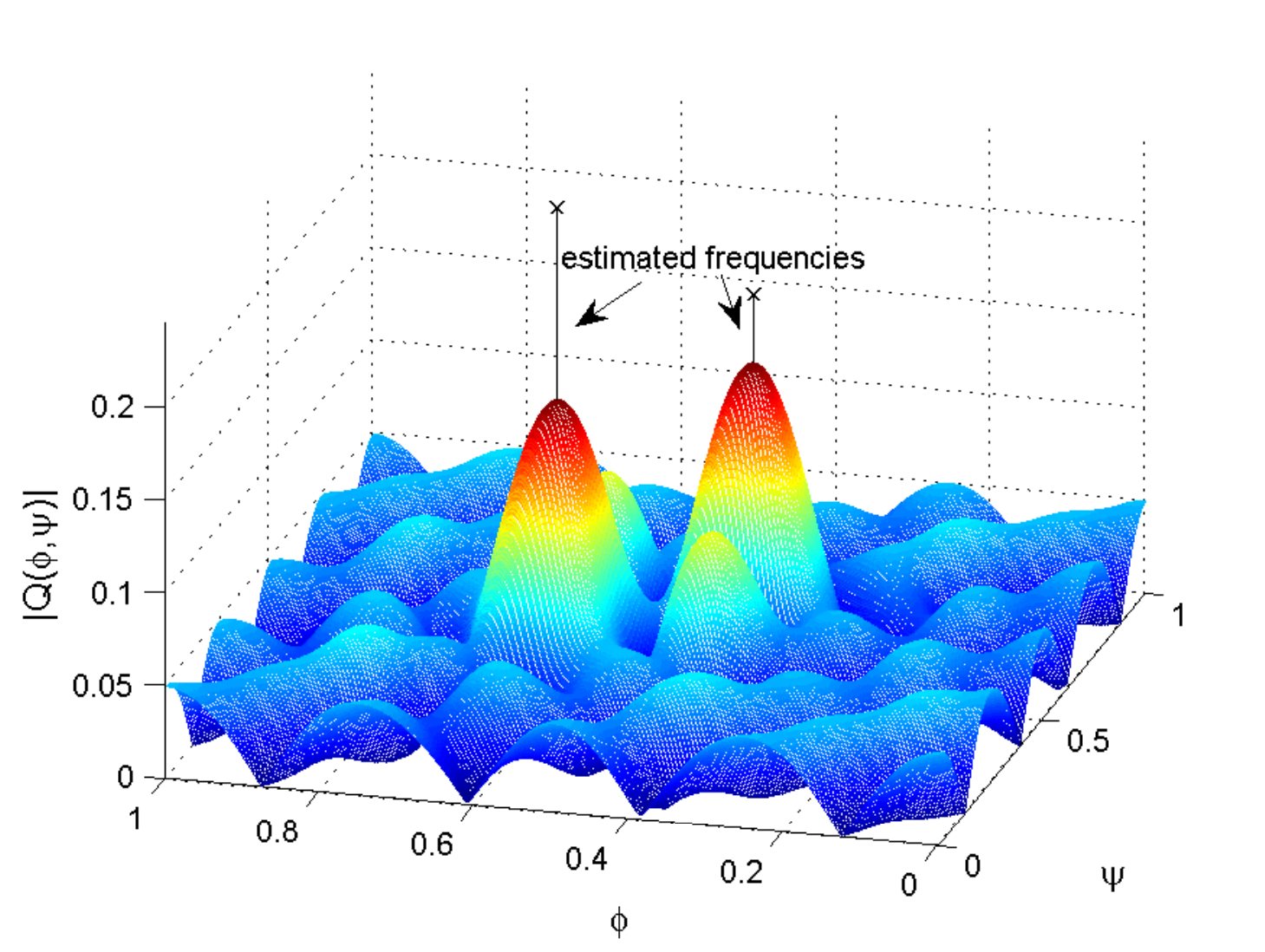}
	
	\caption{Plots of dual polynomial and the estimated frequencies.}
	\label{fig:example3}
\end{figure*}

We give an example of the frequency estimation and demodulation error detection through the dual solution. The parameters for the simulation are $M = 8$, $N = 8$ and $K = 2$. For better visualization, the data carriers are modulated with BPSK. As can be seen from Fig. \ref{fig:example2}a, there are 2 demodulation errors so $\|\bm e \|_0 = 2$. The weighting parameters in \eqref{eq:SDP} are $\lambda = 0.16$ and $\mu = 0.02$. The dual solution is also returned by the solver. According to Fig. \ref{fig:example2}b, the demodulation error can be correctly detected and we have $| \hat \nu_j | = \mu$ when $\hat e_j \ne 0$. In Fig. \ref{fig:example3}, the dual polynomial is plotted. It can be seen that the estimated frequencies are obtained by identifying points where the dual polynomial has magnitude $\lambda$. 

With the estimate of the frequencies, the dictionary matrices can be constructed as 
\begin{eqnarray}
\bm{C}(\bm{\hat \phi},\bm{\hat \psi}) = [\bm a(\hat \phi_1,\hat \psi_1),\bm a(\hat \phi_2,\hat \psi_2),...,\bm a(\hat \phi_{\hat K},\hat \psi_{\hat K})],
\end{eqnarray}
where $\bm{\hat \phi} = \left[ \hat \phi_1, \hat \phi_2,..., \hat \phi_{\hat K} \right]$ and $\bm{\hat \psi} = \left[ \hat \psi_1, \hat \psi_2,..., \hat \psi_{\hat K} \right]$. On this basis, the reflection of the targets and clutters can be estimated using the least-squares (LS) method, i.e.,
\begin{eqnarray}
\label{eq:LS}
\bm{\hat \alpha} = \arg\min_{\bm{\alpha}} \| \bm{\bar r} - \bm{\tilde S}\bm{C}(\bm{\hat \phi},\bm{\hat \psi}) \bm{\alpha} - \bm{\hat e}\|_2^2.
\end{eqnarray}

\section{An ADMM-based Algorithm}

According to the previous section, our high-resolution receiver formulation \eqref{eq:atomicnorm} is equivalent to the SDP given by \eqref{eq:SDP} which can be solved by off-the-shelf solvers such as SeDuMi \cite{sturm1999using} and SDPT3 \cite{toh1999sdpt3}. However, these solvers tend to be slow, especially for large problems. To meet the requirement of real-time signal processing, we next derive a fast method for solving this SDP via the alternating direction method of multipliers (ADMM) \cite{boyd2011distributed}. To put our problem in an appropriate form for ADMM, rewrite \eqref{eq:SDP} as
\begin{eqnarray}
\label{eq:sdp2}
\mathop {\arg \min }\limits_{\bm{\bar z}, \bm{\bar e}, \bm U, t} && \frac{1}{2}{\left\| {{\bm{\bar r}} - \bm{\bar e} - \bm{\tilde S} \bm{\bar z}} \right\|_2^{2}} + \frac{\lambda}{2MN} {\text{Tr}}\left( {\cal T}(\bm U) \right) + \frac{\lambda t}{2} + \mu \|\bm{\bar e}\|_1 + \mathbb{I}_\infty(\bm \Theta \succeq 0) \\
\text{s.t.} && \bm \Theta = \left[ {\begin{array}{*{20}{c}}
		{{\cal T}(\bm U)}& \bm{\bar z} \\
		{{\bm{\bar z}^H}}& t
	\end{array}} \right], \nonumber 
\end{eqnarray}
where $\mathbb{I}_\infty(\cdot)$ is an indicator function that is 0 if the condition in the bracket is true, and infinity otherwise; $\bm U$ is a $(2M-1) \times (2N-1)$ matrix defined by \eqref{eq:U} and \eqref{eq:u}. Dualize the equality constraint via an augmented Lagrangian, we have
\begin{eqnarray}
\label{eq:AL}
{\cal L}_\rho(t,\bm U,\bm{\bar z},\bm{\bar e},\bm{\Upsilon},\bm \Theta) &=&\frac{\lambda}{2} \left( u_0(0) + t \right) +\mu \|\bm{\bar e}\|_1 + \frac{1}{2} \|\bm{\bar r}-\bm{\bar e} - \bm{\tilde S} \bm{\bar z} \|_2^2 + \mathbb{I}_\infty(\bm \Theta \succeq 0) \nonumber \\
&& + \left\langle {{\bm \Upsilon},\bm \Theta  - \left[ {\begin{array}{*{20}{c}}
		{{\cal T}(\bm U)}& \bm{\bar z}\\
		\bm{\bar z}^H & t
		\end{array}} \right]} \right\rangle  
+ \frac{\rho}{2} \left \| {\bm \Theta  - \left[ {\begin{array}{*{20}{c}}
		{\cal T}(\bm U) & \bm{\bar z}\\
		\bm{\bar z} & t
		\end{array}} \right] }\right\|_F^2,
\end{eqnarray}
where $\bm \Upsilon$ is the dual variable, $\langle \bm \Upsilon, \bm \Theta \rangle \triangleq \text{Re} \left( \text{Tr}(\bm \Theta^H \bm \Upsilon) \right)$, $\rho > 0$ is the penalty parameter. The ADMM consists of the following update steps \cite{boyd2011distributed}:
\begin{eqnarray}
\label{eq:UPDA}
(t^{l+1},\bm{\bar z}^{l+1},\bm{\bar e}^{l+1},\bm U^{l+1}) &=& \arg\min_{t,\bm{\bar z},\bm{\bar e},\bm U}{\cal L}_{\rho}(t,\bm U,\bm{\bar z},\bm{\bar e},\bm \Upsilon^l,\bm \Theta^l),\\
\label{eq:UPDA2}
\bm \Theta^{l+1} &=& \arg\min_{\bm \Theta \succeq 0}{\cal L}_{\rho}(t^{l+1},\bm U^{l+1},\bm{\bar z}^{l+1},\bm{\bar e}^{l+1},\bm \Upsilon^l,\bm \Theta),\\
\label{eq:UPDAUpsilon}
\bm \Upsilon^{l+1} &=& \bm \Upsilon^l + \rho \left( {\bm \Theta^{l+1}  - \left[ {\begin{array}{*{20}{c}}
		{{\cal T}(\bm U^{l+1})}& \bm{\bar z}^{l+1}\\
		(\bm{\bar z}^{l+1})^H&{{t^{l+1}}}
		\end{array}} \right] } \right).
\end{eqnarray}

Now we derive the updates of \eqref{eq:UPDA} and \eqref{eq:UPDA2} in detail. For convenience, the following partitions are introduced:
\begin{eqnarray}
\bm \Theta^l = \left[ {\begin{array}{*{20}{c}}
	{\bm \Theta _{0}^l}&{\bm \theta _{1}^l}\\
	{(\bm \theta_{1}^l)^H}&{\bar \Theta^l}
	\end{array}} \right],
\end{eqnarray}
\begin{eqnarray}
\label{eq:Upsilon}
\bm \Upsilon^l = \left[ {\begin{array}{*{20}{c}}
	{\bm \Upsilon_0^l}&{\bm \upsilon_1^l}\\
	{(\bm \upsilon_1^l)^H}&{\bar \Upsilon^l}
	\end{array}} \right],
\end{eqnarray}
where $\bm \Theta_0^l$ and $\bm \Upsilon_0^l$ are $MN \times MN$ matrices, $\bm \psi_1^l$ and $\bm \upsilon_{1}^l$ are $MN \times 1$ vectors, $\bar \Theta^l$ and $\bar \Upsilon^l$ are scalars. Computing the derivative of ${\cal L}_\rho(t,\bm U,\bm{\bar z},\bm{\bar e},\bm{\Upsilon},\bm \Theta)$ with respect to $\bm{\bar z}$, $t$ and the elements of $\bm U$, we have
\begin{eqnarray}
\nabla_{\bm{\bar z}} {\cal L}_{\rho} &=& \bm{\tilde S}^H(\bm{\tilde S}\bm{\bar z} + \bm{\bar e}^l - \bm{\bar r}) - 2 \bm \upsilon_1^l + 2\rho (\bm{\bar z} - \bm \theta_{1}^l),\\
\nabla_{t} {\cal L}_{\rho} &=& \frac{{{\lambda}}}{2} - \bar\Upsilon^l + \rho (t - {\bar\Theta^l}),\\
\nabla_{u_j(k)}{\cal L}_{\rho} &=& \left\{ \begin{array}{l}
\frac{{{\lambda}}}{2} + N\rho u_j(k) - {\rm{Tr}}(\rho \bm \Theta _0^l + \bm \Upsilon _0^l),j = k = 0,\\
(N - j)(M - k)\rho u_j(k) - \sum_{m=0}^{N-j-1}{\rm{Tr}}_{k}({\cal S}_{j,k}(\rho \bm \Theta_{0}^l + \bm \Upsilon _{0}^l)),j \ne 0 \text{ or } k \ne 0.
\end{array} \right.
\end{eqnarray}
For $\bm P \in \mathbb{C}^{MN \times MN}$, ${\cal S}_{j,k}(\bm P)$ returns the $(j,k)$-th $M \times M$ submatrix $\bm P_{j,k}$. $\text{Tr}_{k}(\cdot)$ outputs the trace of the $k$-th sub-diagnal of the input matrix. $\text{Tr}_{0}(\cdot)$ outputs the trace of the input matrix.

By setting the derivatives to 0, $\bm{\bar z}^{l+1}$, $t^{l+1}$ and $\bm u^{l+1}$ can be updated by:
\begin{eqnarray}
\label{eq:UPDAx1}
\bm{\bar z}^{l+1} &=& (\bm{\tilde S}^H \bm{\tilde S} + 2\rho \bm{E}_N)^{-1}(\bm{\tilde S}\bm{\bar r} - \bm{\tilde S}\bm{\bar e}^{l} + 2\rho \bm \theta_{1}^l + 2 \bm \upsilon_{1}^l),\\
\label{eq:UPDAtheta}
t^{l+1} &=& \bar \Theta^l + \left(\Upsilon^l - \lambda/2 \right)/\rho,\\
\label{eq:UPDAu}
\bm U^{l+1} &=& {\cal T}^*(\bm \Theta_{0}^l + \bm \Upsilon_{0}^l/\rho) - \frac{\lambda}{2 M N \rho} \bm I_1 ,
\end{eqnarray}
where $\bm E_N$ is an $N \times N$ identity matrix, $\bm I_1 = [1,0,0,...,0]^T$, ${\cal T}^*(\cdot)$ denotes the adjoints of the map ${\cal T}(\cdot)$. Suppose $\bm Q = {\cal T}^*(\bm P)$ where $\bm Q = [\bm q_{-N+1},\bm q_{-N+2},...,\bm q_{N-1}]$ with $\bm q_j=[q_j(-M+1),q_j(-M+2),...,q_j(M-1)]^T$. Then we have
\begin{eqnarray}
\label{eq:adjoint}
q_j (k) = \frac{1}{(N-j)(M-k)}\sum_{m=0}^{N-j-1} \text{Tr}_{k} ({\cal S}_{j,m}(\bm P)),
\end{eqnarray}
for $j= -N+1,-N+2,...,N-1$ and $k= -M+1,-M+2,...,M-1$. 

The term in \eqref{eq:AL} that is related to $\bm{\bar e}$ is $\mu \|\bm{\bar e}\|_1 + \frac{1}{2} \|\bm{\bar r}-\bm{\bar z}^{l} - \bm{\bar e}\|_2^2$. Hence, $\bm{\bar e}$ can be updated by 
\begin{eqnarray}
\label{eq:updatee}
\bm{\bar e}^{l+1}=\arg \min_{\bm{\bar e}} \mu \|\bm{\bar e}\|_1 + \frac{1}{2} \|\bm{\bar r} - \bm{\tilde S} \bm{\bar z}^{l}-\bm{\bar e}\|_2^2,
\end{eqnarray}
which can be easily achieved by the proximal operator \cite{maleki2013asymptotic}:
\begin{eqnarray}
\bm{\bar e}^{l+1}=\text{prox}_{\mu}(\bm{\bar r}- \bm{\tilde S}\bm{\bar z}^{l}),
\end{eqnarray}
where $\text{prox}_{\mu}(\bm{\bar e}) = [\text{prox}_{\mu}(\bar e_1), \text{prox}_{\mu}(\bar e_2),...,\text{prox}_{\mu}(\bar e_{MN})]^T$ with
\begin{eqnarray}
\label{eq:prox}
\text{prox}_{\mu}(\bar e_i) = (|\bar e_i|-\mu) \text{sign}(\bar e_i) \mathbb{I}_0(|\bar e_i|>\mu),
\end{eqnarray}	
where $\mathbb{I}_0(\cdot)$ is an indicator function that goes to 0 if the condition in the bracket is not satisfied, and 1 otherwise. The update of $\bm \Theta^l$ is simply the projection onto the positive semidefinite cone
\begin{eqnarray}
\label{eq:UPDAPhi}
\bm \Theta^{l+1} = \arg\min_{\bm \Theta \succeq 0} \left \| {\bm \Theta  - \left[ {\begin{array}{*{20}{c}}
		{{\cal T}(\bm U^{l+1})}& \bm{\bar z}^{l+1}\\
		(\bm{\bar z}^{l+1})^H&{{t^{l+1}}}
		\end{array}} \right] } + \bm \Upsilon^{l+1}/\rho \right\|_F^2,
\end{eqnarray}
where projecting a matrix onto the positive definite cone is accomplished by forming an eigenvalue decomposition of the matrix and setting all negative eigenvalues to zero. Noticing that the submatrix $\bm{\hat \upsilon}_1$ in $\bm{\hat \Upsilon}$ corresponds to the dual variables, ADMM also provides the dual solution to \eqref{eq:dual}. The following lemma can be used to obtain the dual solution $\bm{\hat \nu}$.

\begin{lemma}
	Suppose the solution to \eqref{eq:dual} is $\bm{\hat \nu}$, and the dual solution to \eqref{eq:sdp2} is $\bm{\hat \Upsilon}$, then $\bm{\hat \nu} = -\frac{\bm{\hat \upsilon_1}}{2}$ where $\bm{\hat \upsilon}_1$ is given by \eqref{eq:Upsilon} with $\bm \Upsilon$ and $\bm{\upsilon}_1$ replaced by $\bm{\hat \Upsilon}$ and $\bm{\hat \upsilon}_1$ respectively.
\end{lemma}

The proof of Lemma 2 is given in Appendix A. For clarity, we summarize the proposed CS-ANL1 algorithm in Table I. Note that when there is no demodulation error, the CS-AN algorithm is also given by Table I with $\bm{\bar e}=\bm 0$ over the iterations. 
\begin{table}[htbp]
	\label{tab:A1}
	\caption{CS-ANL1 Algorithm}
	\begin{tabular}{lcl}
		\toprule
		Input $\bm{\bar r}$, $M$, $N$, $L$\\
		\midrule
		Initialize $\bm{\bar e}^0 = \bm 0$, $\bm{\bar z}^0 = \bm 0$, $\bm U = \bm 0$, $\bm \Theta^0 = \bm 0$, $\bm \Upsilon^0 = \bm 0$, $t^0 = 0$.\\
		\sf{For $l=1$ to $L$ } \\
		\hspace{0.4cm} 1, Update $\bm{\bar z}^{l+1}$, $t^{l+1}$, ${\bm U}^{l+1}$ and $\bm{\bar e}^{l+1}$ according to \eqref{eq:UPDAx1}-\eqref{eq:updatee} \\
		\hspace{0.4cm} 2, Update $\bm \Theta^{l+1}$ according to \eqref{eq:UPDAPhi} \\
		\hspace{0.4cm} 3, Update $\bm{\Upsilon}^{l+1}$ according to \eqref{eq:UPDAUpsilon}. \\
		\sf{End for} \\
		\hspace{0.4cm} 4, Obtain the dual solution $\bm{\hat \nu} = -\frac{\bm{\hat \upsilon_1}}{2}$ with $\bm{\hat \upsilon}_1$ obtained from \eqref{eq:Upsilon}. \\
		\hspace{0.4cm} 5, Identify $(\bm{\hat \phi},\bm{\hat \psi})$ by solving $\lambda-|\langle \bm{\hat \nu}, \bm a(\phi,\psi) \rangle|^2=0$ where $\phi$ \\
		\hspace{0.8cm} and $\psi$ are searched over a set of grids.\\
		\hspace{0.4cm} 6, Estimate the amplitude $\bm{\hat \alpha}$ via \eqref{eq:LS}.\\
		\midrule
		Output $\bm{\hat \alpha}$, $\bm{\hat \psi}$, $\bm{\hat \phi}$. \\
		\bottomrule
	\end{tabular}
\end{table}

\section{Simulation Results}
\subsection{Simulation Setup}
\label{sec:setup}

To demonstrate the performance of the proposed algorithm, we simulate a scenario with multiple targets and clutters. There is one illuminator which transmits OFDM signal with a carrier frequency of 2GHz. The frequency spacing of the adjacent subcarrier is 5kHz. The number of subcarrier frequencies is $N=64$, so the bandwidth of the communication signal is 320kHz. The symbol length is $T = 200 \mu s$ and the cyclic prefix is $T_\text{cp} = T/2$, so the block length is $\bar T = 300 \mu s$. We use $M = 16$ blocks of data for signal processing, so the time of data collection is $4.8ms$.

Both the targets and clutters are assumed to be point scatterers in our simulations. The power of one target is set as -40dB per sample, and the other two targets are set as -50dB per sample. The amplitudes of the clutters are randomly generated according to a Gaussian distribution, and the variance is controlled by the power of the clutters. The powers of the direct path and clutters at each sampling instant are set as 0dB and -10dB, respectively. The power of the noise is set as -40dB. Hence, for the strong target, the SNR and SCR are 0dB and -30dB respectively. For the weak targets, the SNR and SCR are -10dB and -40dB respectively.

\begin{figure*}
	\centering
	\subfloat[][]{\includegraphics[width=3.2in]{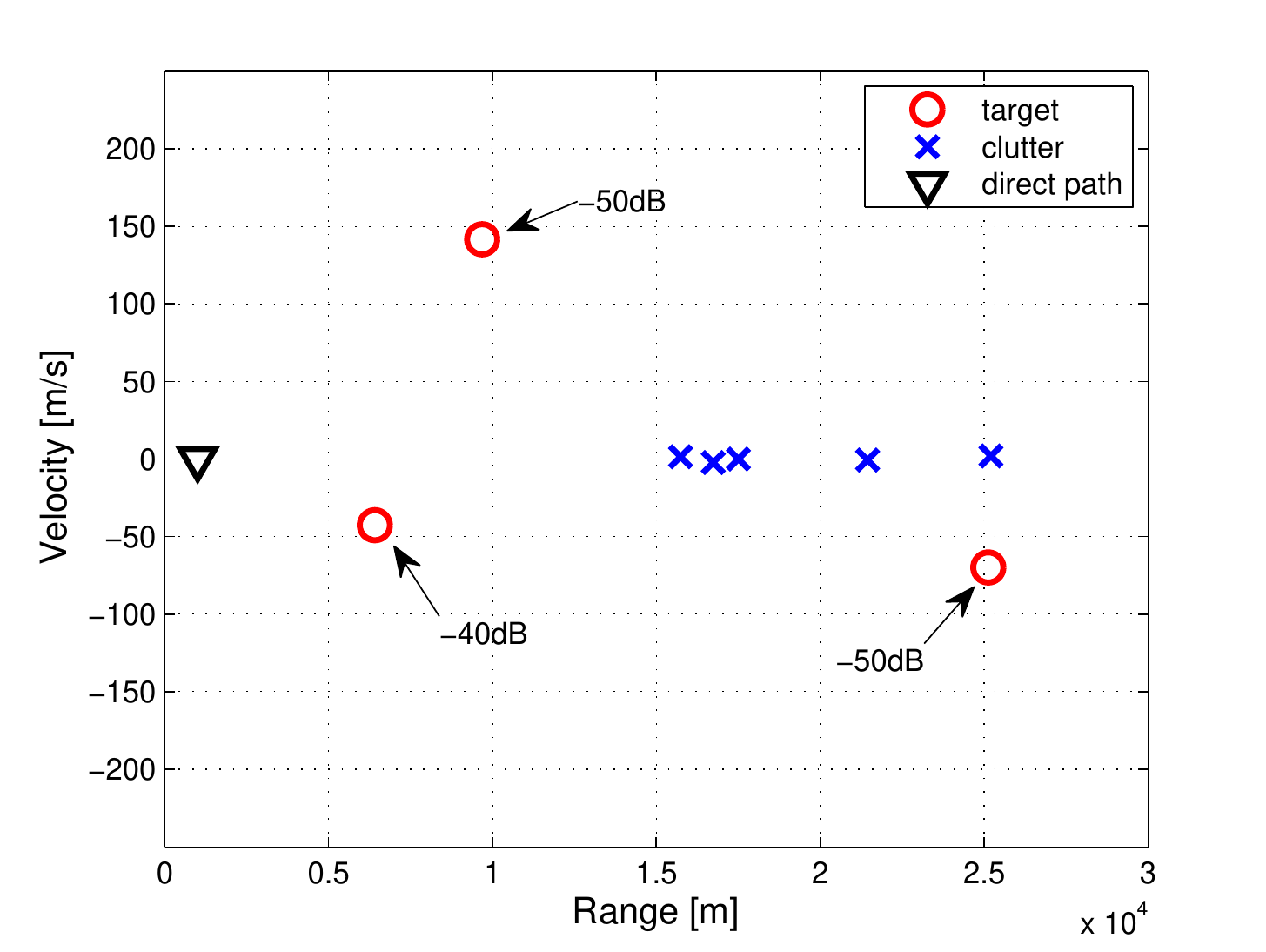}}
	\subfloat[][]{\includegraphics[width=3.2in]{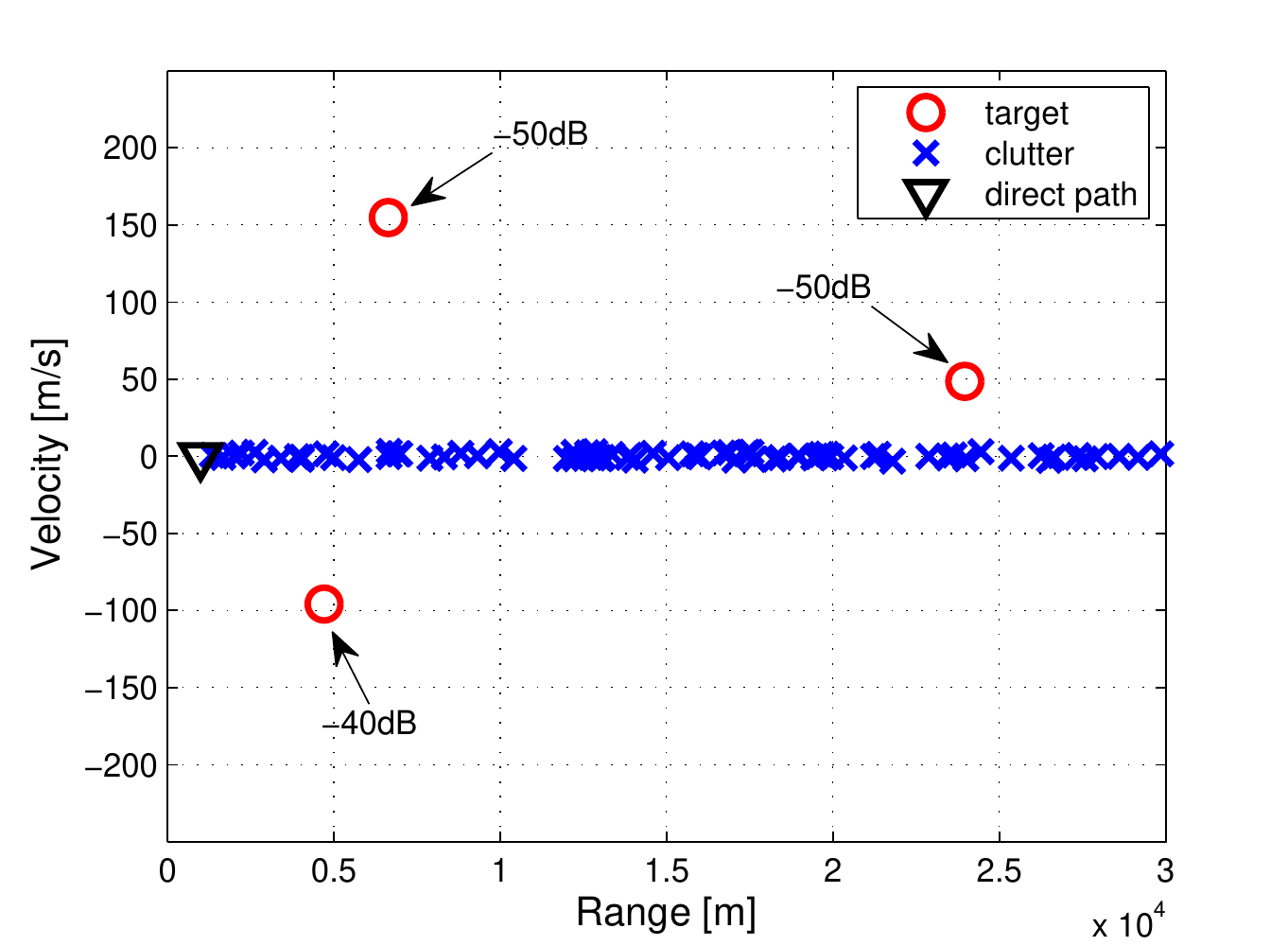}}
	
	\caption{Simulation scenarios: (a) scenario 1 with 5 clutters and 3 targets; (b) scenarios 2 with 80 clutters and 3 targets.}
	\label{fig:scenario}
\end{figure*}

The algorithms are tested for two scenarios shown in Fig. \ref{fig:scenario} where the range and velocities of each target or clutter is given. As can be seen from the figure, there are three targets in both scenarios, but the number of clutters are different. In the first scenario, the number of clutters is 5, corresponding to the low clutter density. In the second scenario, the number of clutters is 80, corresponding to the high clutter density in low Doppler frequency. Therefore, with respect to \eqref{eq:rk3}, we have $K = 9$ for Scenario 1 and $K=84$ for Scenario 2.

We will compare the performance of the proposed algorithm with that of the super-resolution receivers based on CS-L1 and 2D-MUSIC. The delays and Doppler frequencies are estimated based on the dual solution $\bm{\hat \nu}$. To get a clear visual comparison, we also plot the 2D-MUSIC spectrum for the proposed CS-AN and CS-ANL1 algorithms. Specifically, after the proposed algorithms outputs $\bm{\hat z} = [\hat z(0), \hat z(1),...,\hat z(MN-1)]^T$, the 2D-MUSIC algorithm is used to obtain the spectrum in the delay-Doppler plane with $R_{m}(n) = \hat z(mM+n)$ as the input. Some other parameters of the simulations are given as follows. 

1, The range of both targets and clutters are uniformly generated in the surveillance area, ranging from 1km to 30km. The velocities of the clutters are randomly generated between -3m/s and 3m/s. The velocities of the targets are randomly generated between -156m/s and 156m/s.

2, The quadrature phase-shift keying (QPSK) modulation is used. In the situation when the demodulation error is considered, the mistaken demodulation is controlled by the bit-error-rate (BER).

3, For the proposed algorithm, the weighting parameters are set as $\lambda = \sigma \sqrt{MN \log(MN)}$ and $\mu = \frac{\lambda}{\sqrt{MN}}$. The weight for the augmented Lagrangian is set as $\rho =0.05$. 

4, For the CS-L1 algorithm, we discretize the continuous parameter space to a finite set of grid points. The number of grids in our simulations is set as $\tilde L = \tilde M \tilde N$ where $\tilde M=4M$ and $\tilde N= 4N$. The weighting parameter for the CS-L1 method is $\gamma = 2\sigma \sqrt{2 \log(\tilde L)}$.

5, The 2D-MUSIC algorithm requires to construct an $(N'M') \times N_\text{snap}$ observation matrix via spatial smoothing. In our simulation, the parameters are set as $M' =M/2$, $N' = N/2$ and $N_\text{snap} = (M/2+1)(N/2+1)$. 

6, With the delay and Doppler frequency, we can compute the range and velocity of the targets. Specifically, if the delay and Doppler frequencies are $(\tau_k, f_k)$, then the range and velocity are $(\tau_k c, f_k c/f_c)$ where $c$ is the speed of light and $f_c$ is the carrier frequency. 

\subsection{Performance}

\begin{figure*}
	\centering
	\subfloat[][]{\includegraphics[width=3.2in]{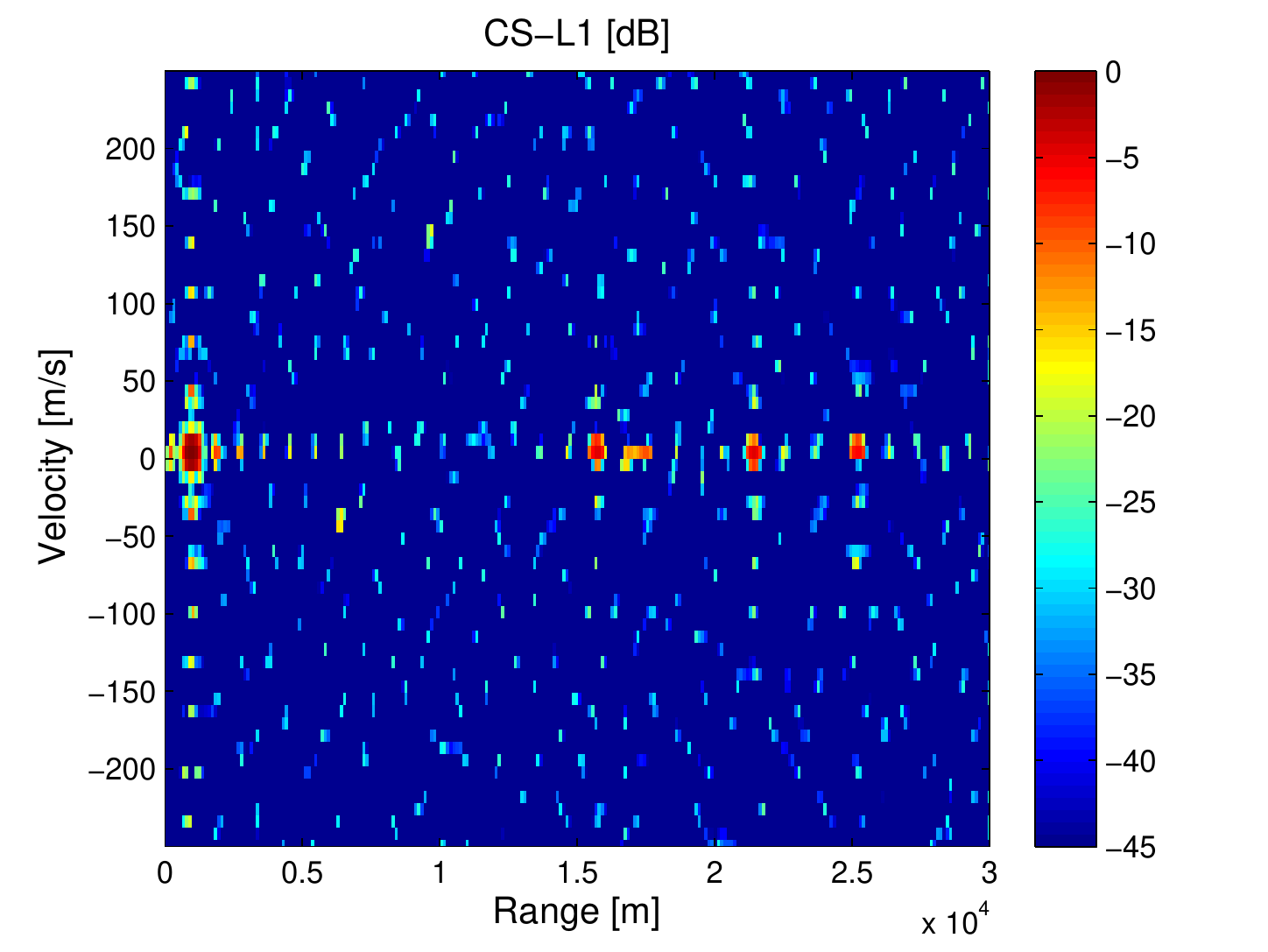}}
	\subfloat[][]{\includegraphics[width=3.2in]{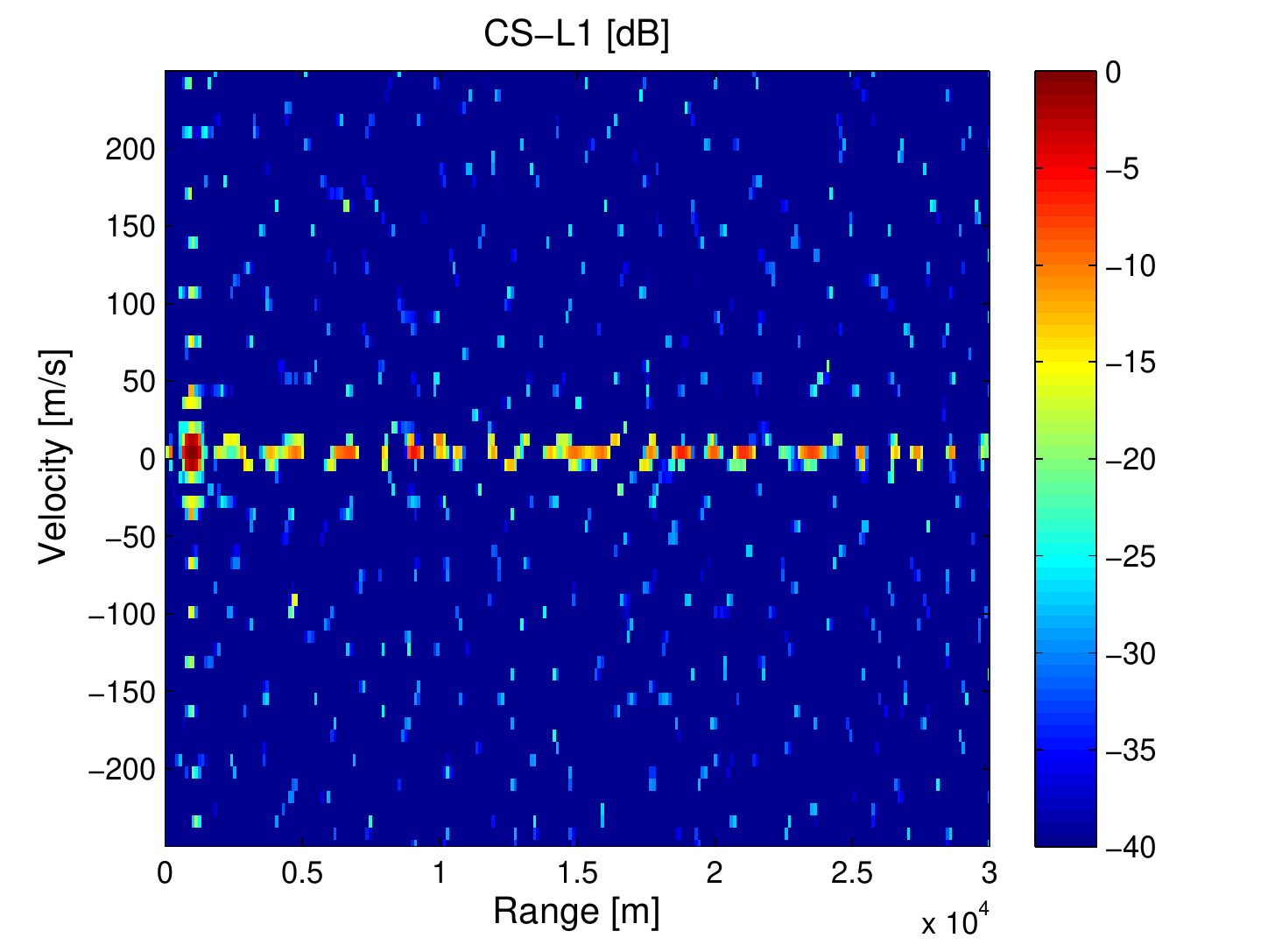}}
	
	\subfloat[][]{\includegraphics[width=3.2in]{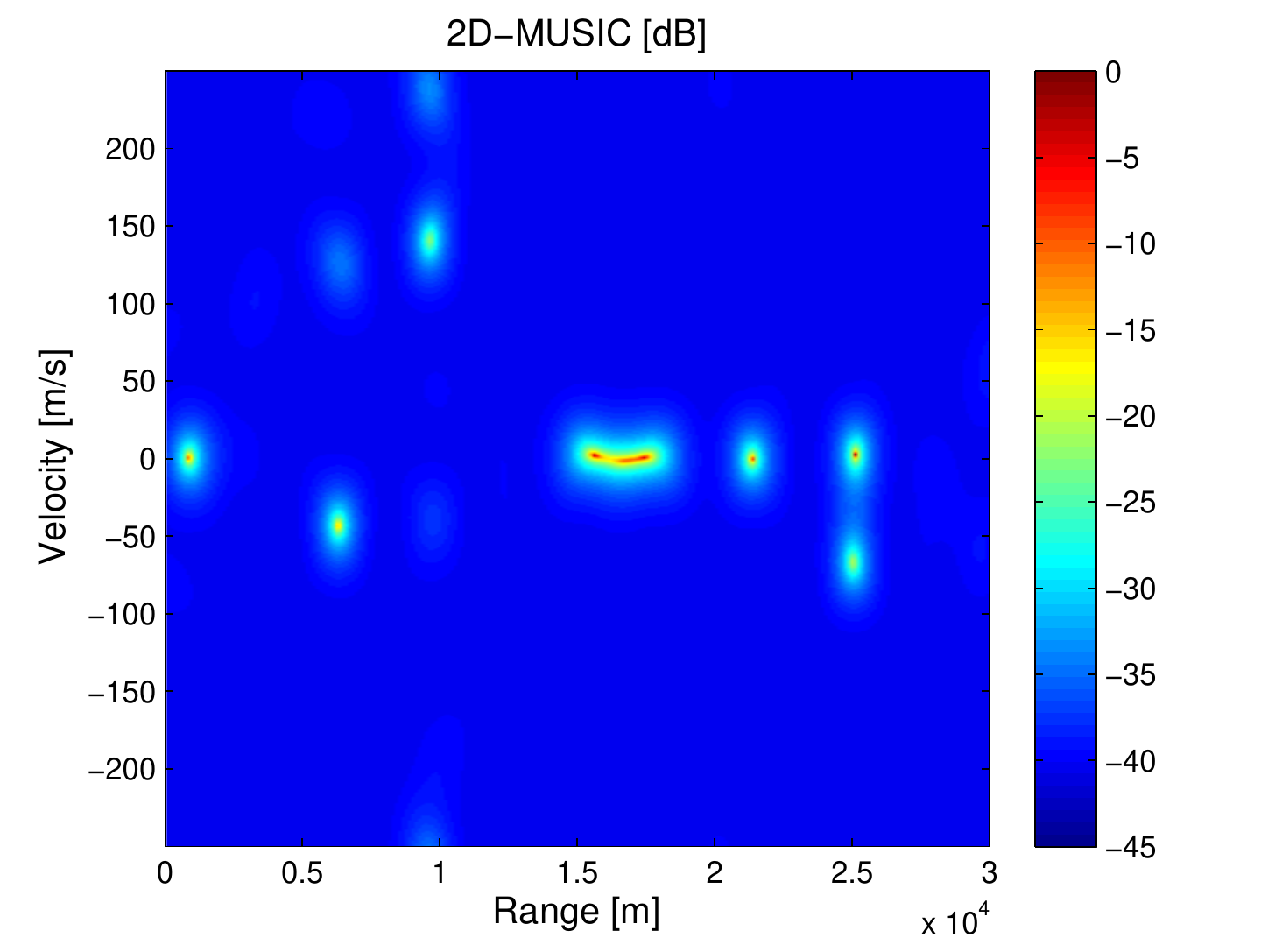}}
	\subfloat[][]{\includegraphics[width=3.2in]{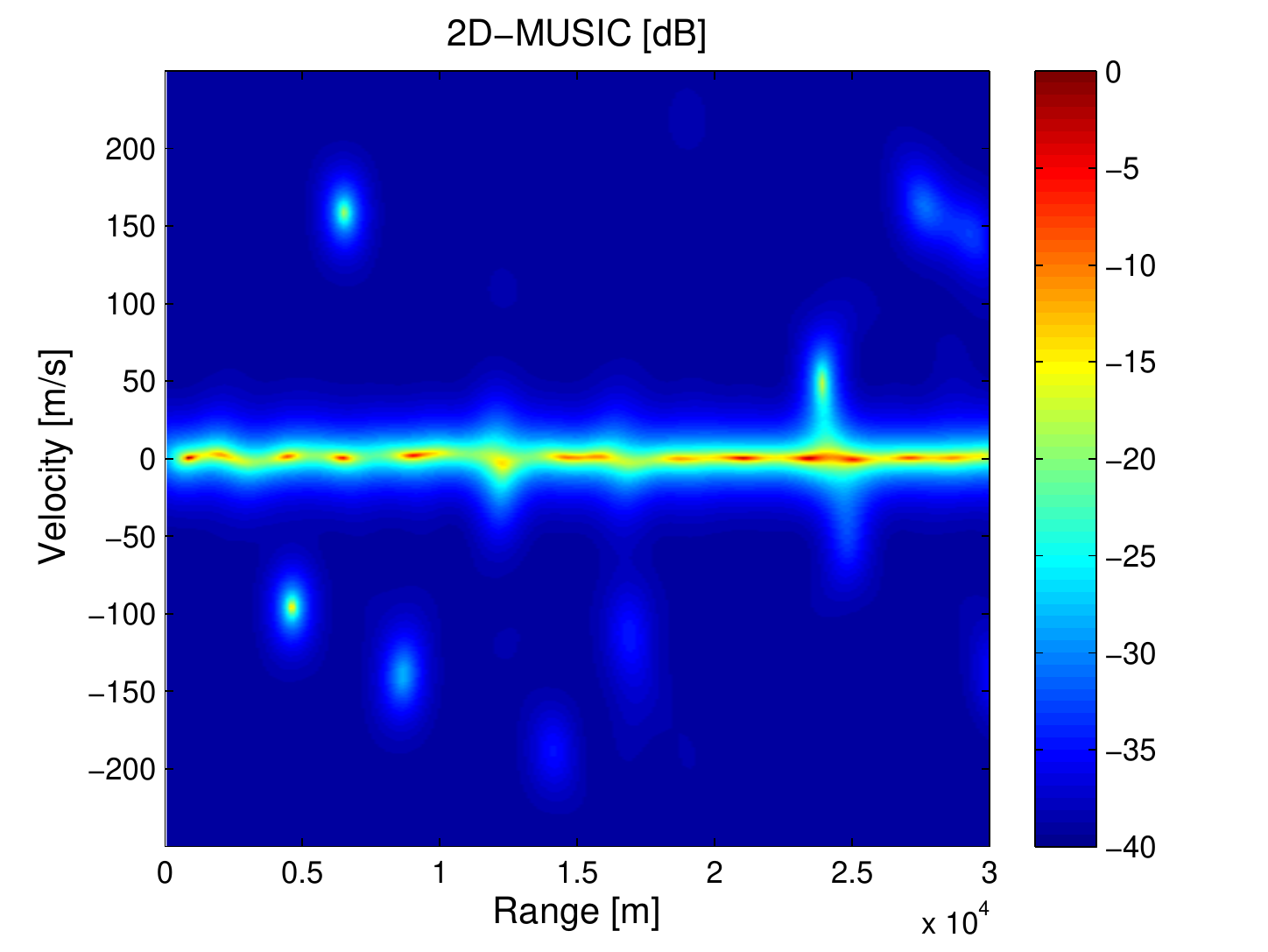}}
	
	\subfloat[][]{\includegraphics[width=3.2in]{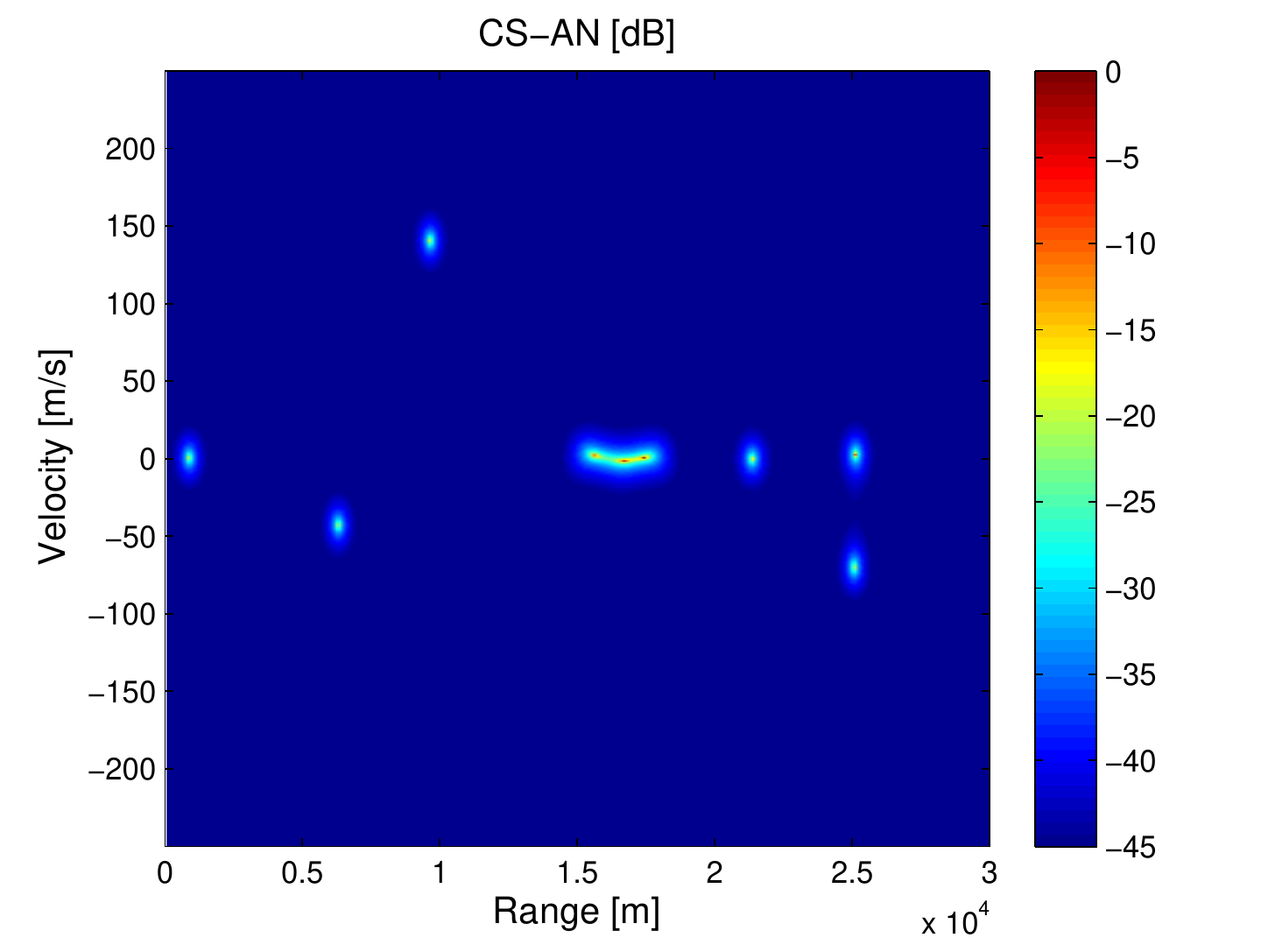}}
	\subfloat[][]{\includegraphics[width=3.2in]{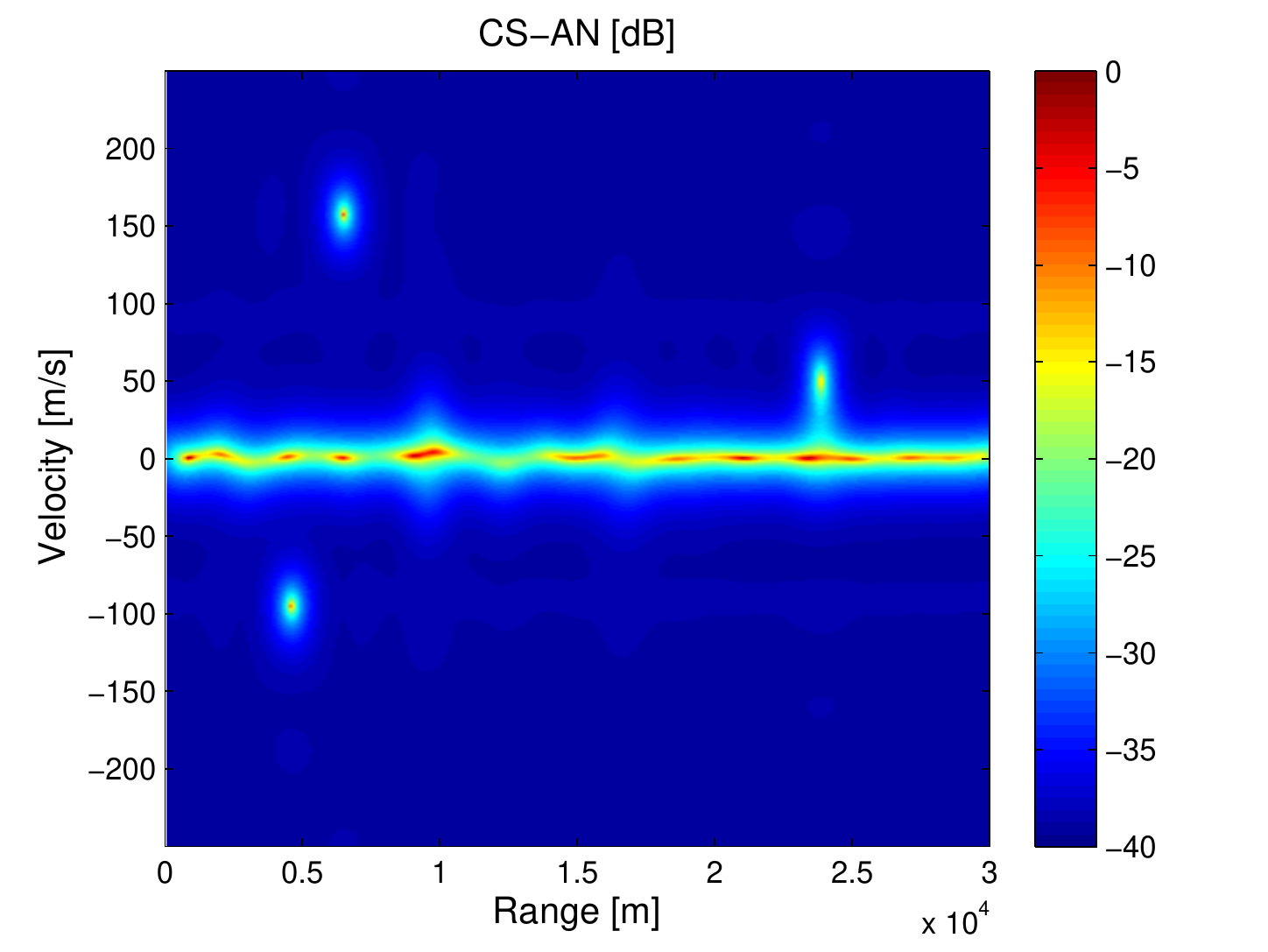}}
	
	\caption{Simulation result of CS-L1, 2D-MUSIC and CS-AN when BER = 0: (a)(c)(e) Scenario 1, and (b)(d)(f) Scenario 2. The positions of the targets in both scenarios are given in Fig. \ref{fig:scenario}.}
	\label{fig:BER0}
\end{figure*}

We firstly compare the CS-L1, 2D-MUSIC and CS-AN algorithms when the communication is error free. As can be seen from Fig. \ref{fig:BER0}, the CS-L1 algorithm will result in more false alarms due to the model mismatch. As the target may not exactly locate on the grid, its energy is split over adjacent grid points, which degrades the energy of the target and the accuracy of localization. Both 2D-MUSIC and CS-AN do not have the off-grid problem, but the spectrum of the CS-AN algorithm contains less ghost peaks, indicating that the CS-AN algorithm has better denoising performance. 

It is worth noting that all the algorithms cannot distinguish the clutters when their separation is too small, especially in Scenario 2 with high clutter density. However, it does not affect the estimation of targets' delays and Doppler frequencies because the moving targets are of higher Doppler frequencies, and the clutters can be easily eliminated by removing the paths around zero velocity region. The targets, however, are usually widely separated and can be distinguished easily.

\begin{figure*}
	\centering
	\subfloat[][]{\includegraphics[width=3.2in]{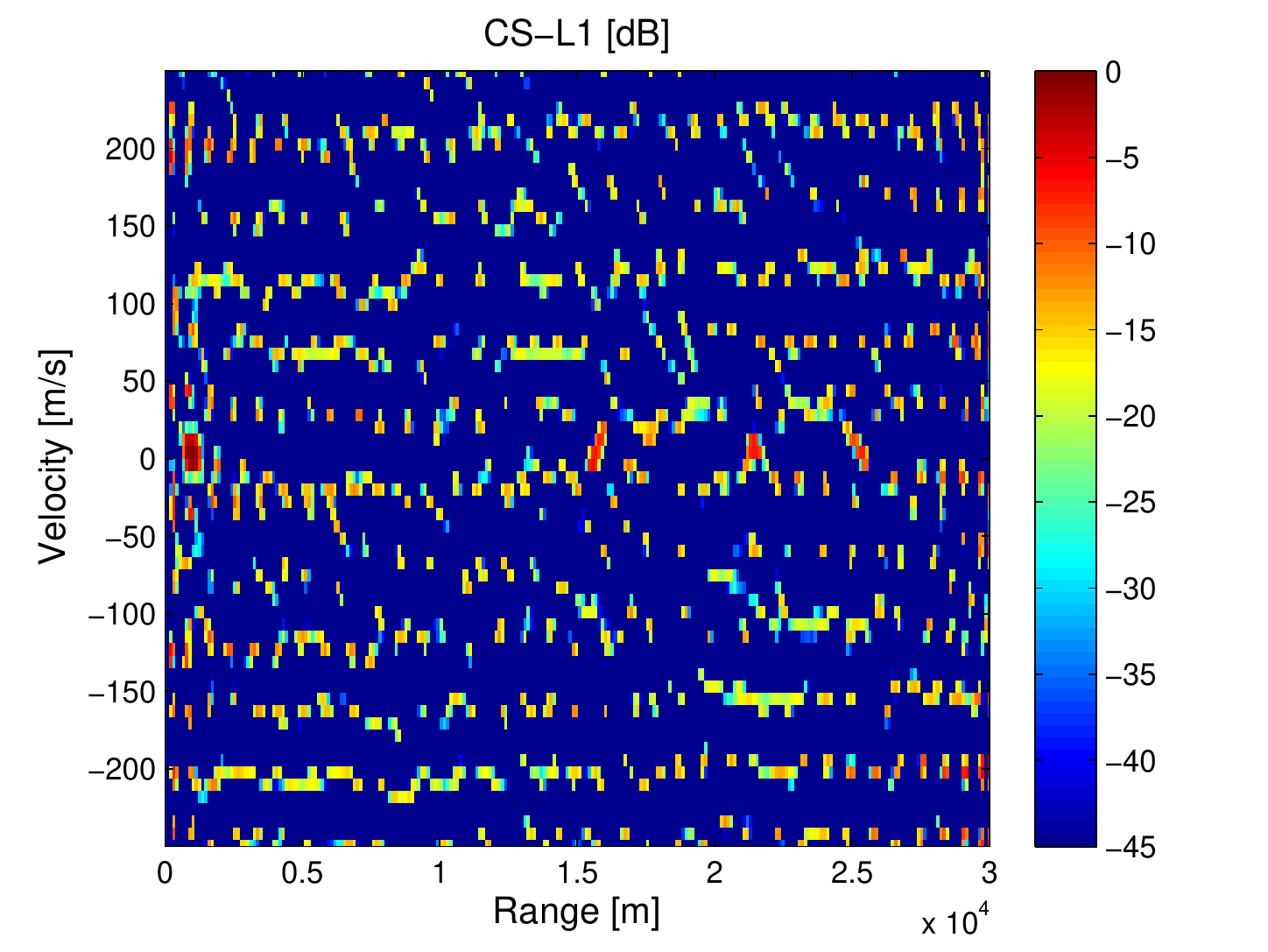}}
	\subfloat[][]{\includegraphics[width=3.2in]{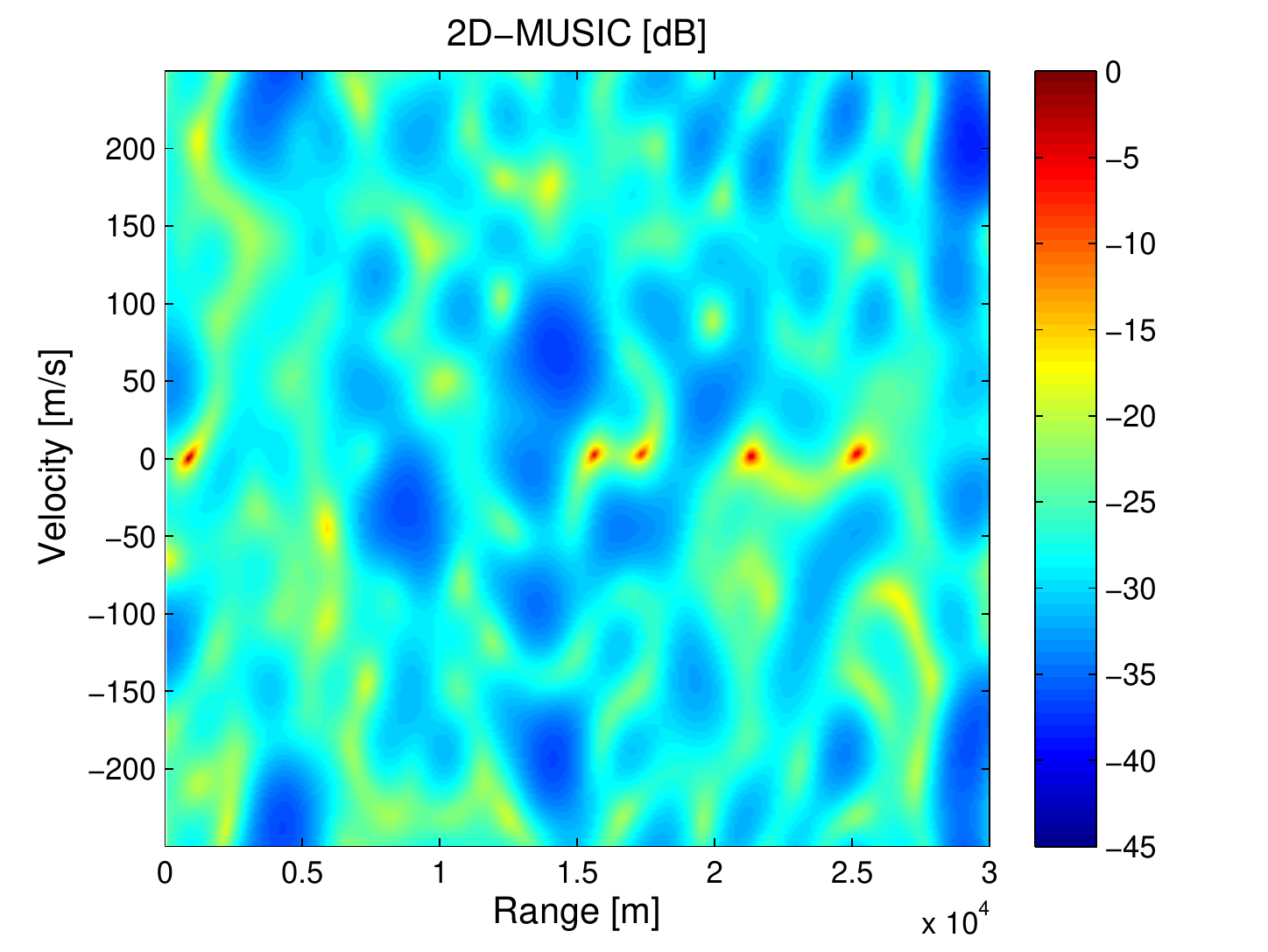}}
	
	\subfloat[][]{\includegraphics[width=3.2in]{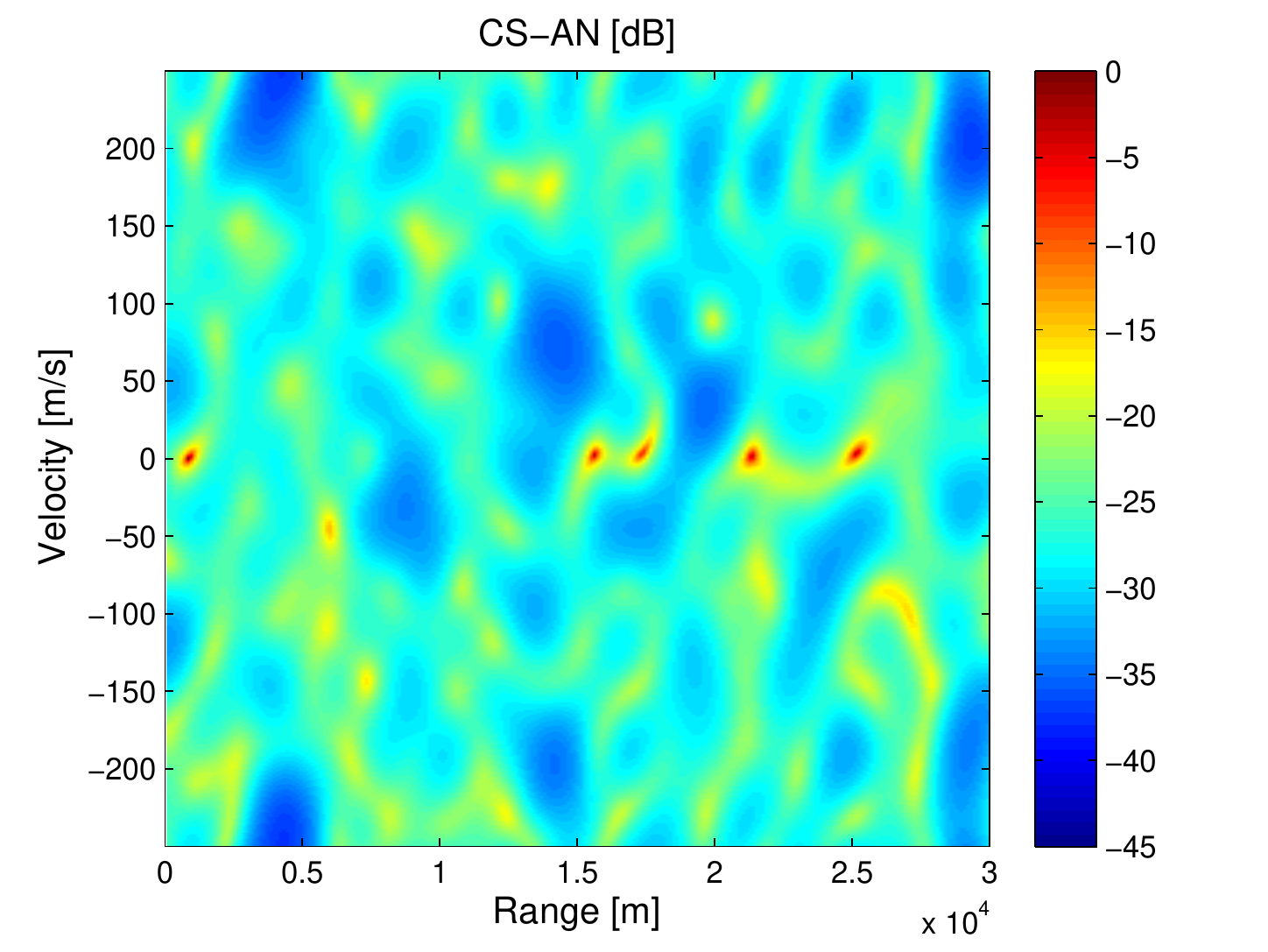}}
	\subfloat[][]{\includegraphics[width=3.2in]{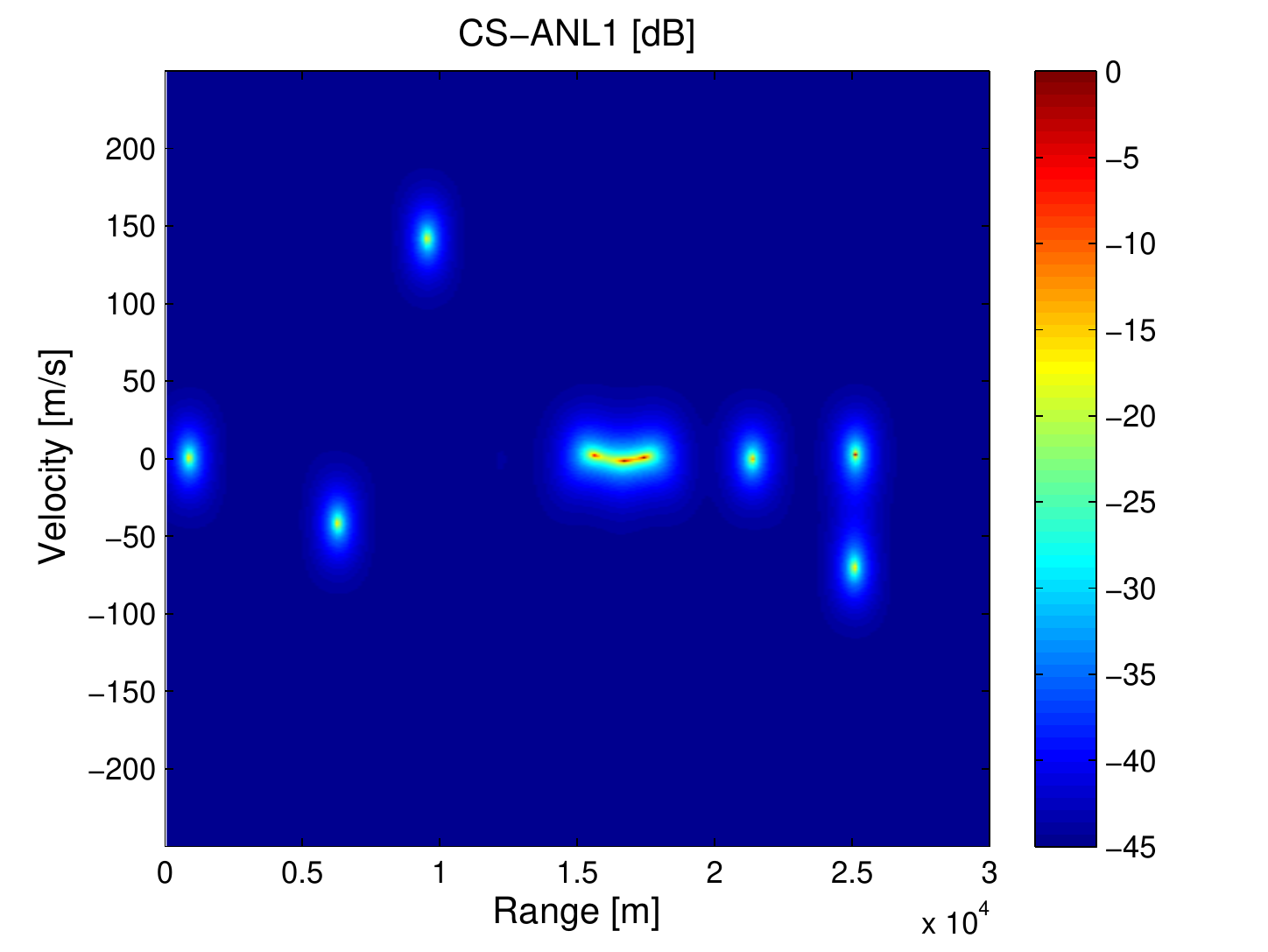}}
	\caption{Simulation results of the algorithms in Scenario 1 when BER=0.02. The positions of the targets are given in Fig. \ref{fig:scenario}a.}
	\label{fig:BER_S1}
\end{figure*}

\begin{figure*}
	\centering
	\subfloat[][]{\includegraphics[width=3.2in]{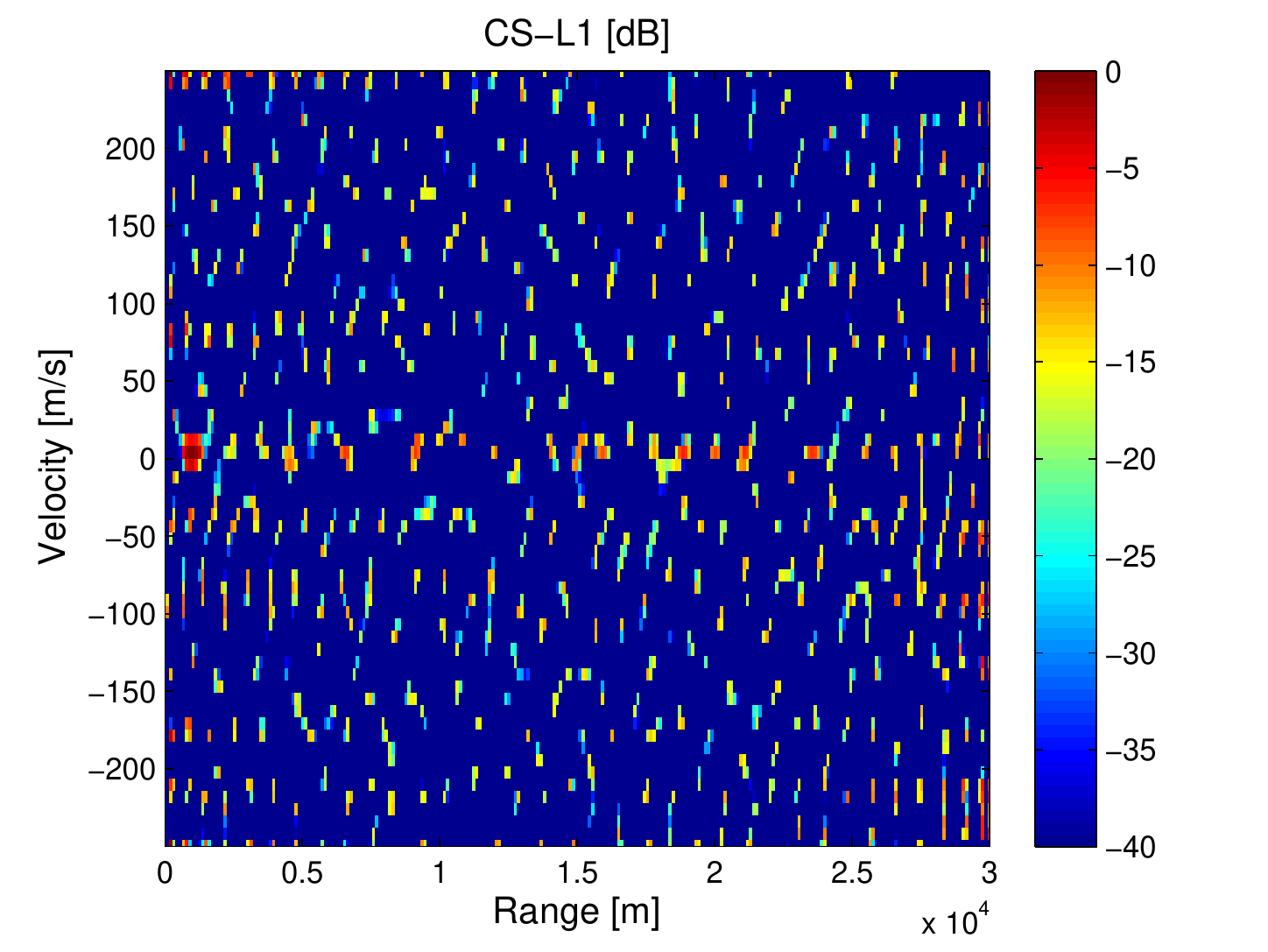}}
	\subfloat[][]{\includegraphics[width=3.2in]{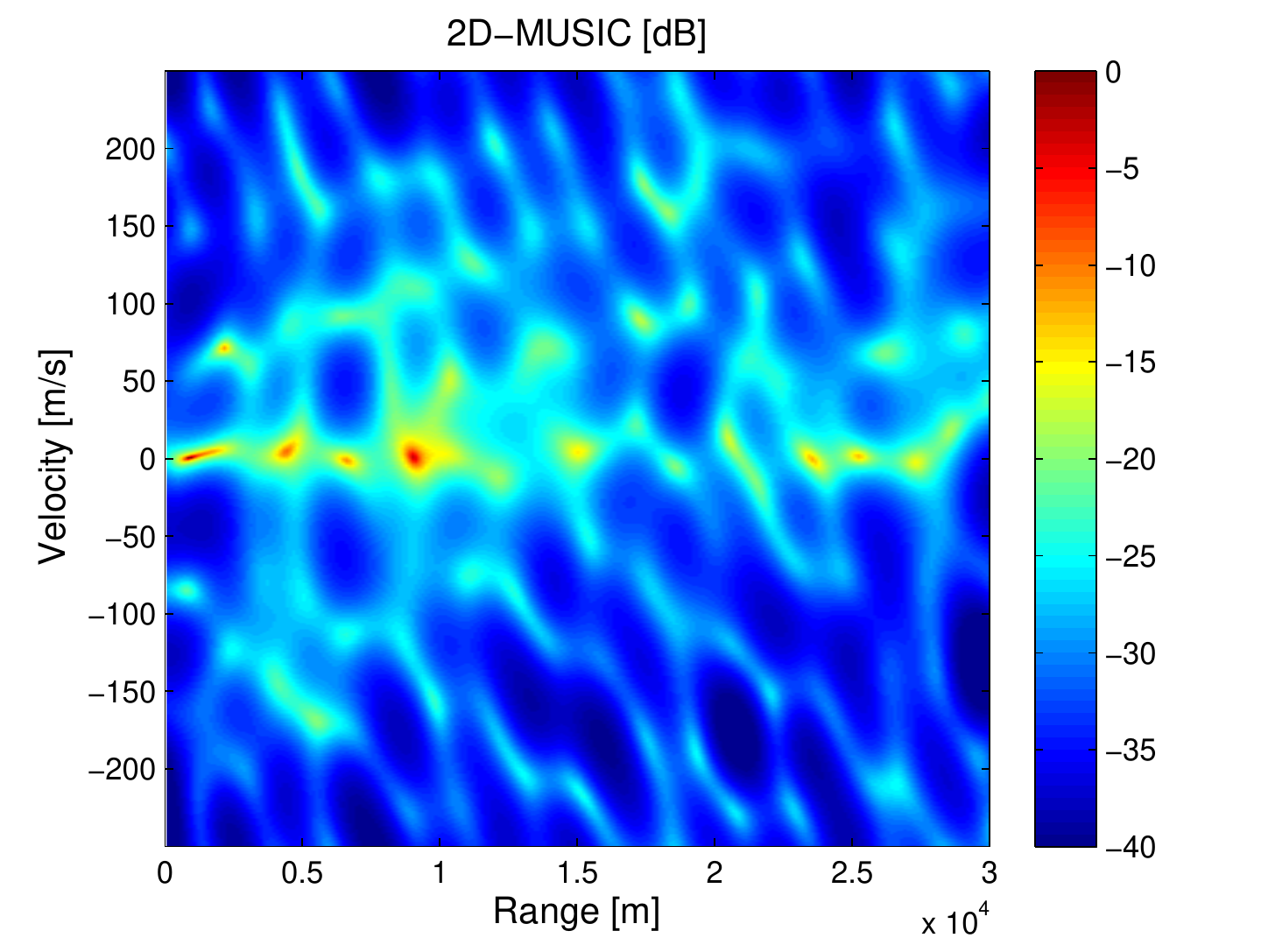}}
	
	\subfloat[][]{\includegraphics[width=3.2in]{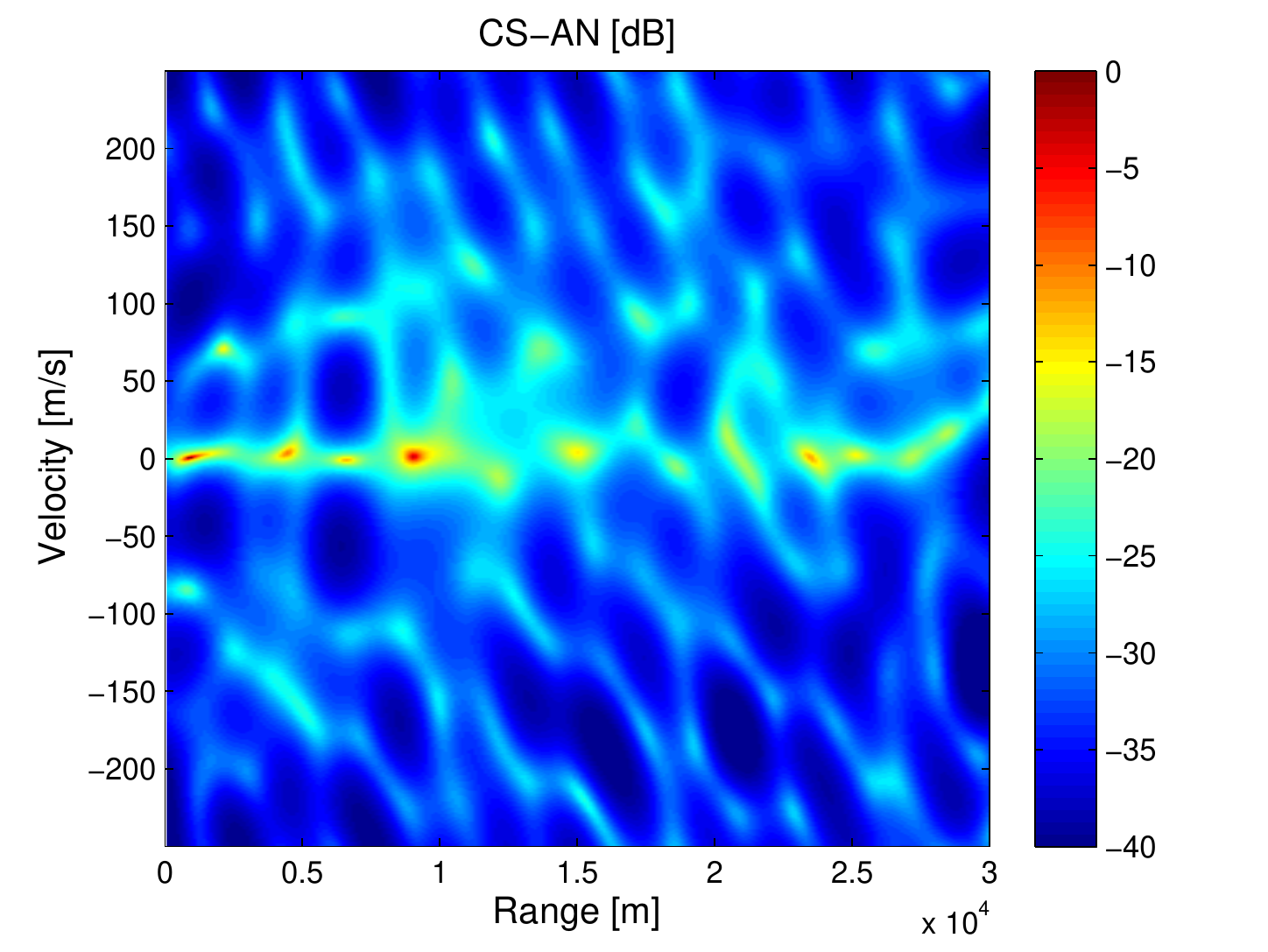}}
	\subfloat[][]{\includegraphics[width=3.2in]{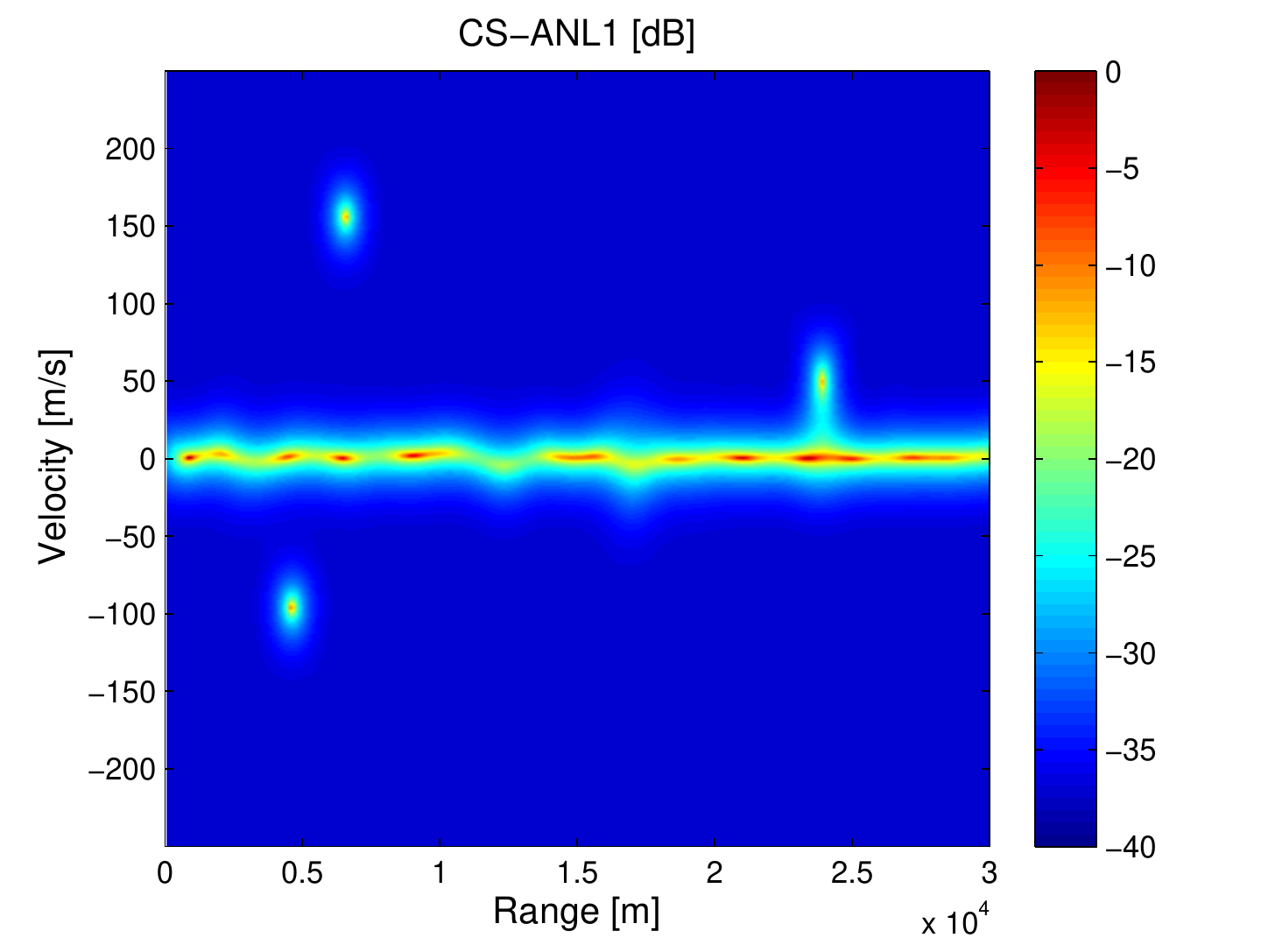}}
	
	\caption{Simulation results of the algorithms in Scenario 2 when BER=0.02. The positions of the targets are given in Fig. \ref{fig:scenario}b.}
	\label{fig:BER_S2}
\end{figure*}

We then illustrate the performance of the proposed algorithm in the presence of demodulation errors. In Fig. \ref{fig:BER_S1} and Fig. \ref{fig:BER_S2}, we compare the algorithms in both scenarios when the BER is 0.02. According to the result, the demodulation errors substantially degrades the performance of the CS-L1, 2D-MUSIC and CS-AN algorithms. For the CS-L1 algorithm, the false alarm rate is high, which causes much interference to the subsequent signal processing. For the 2D-MUSIC and CS-AN algorithms, the mainlobes are defocused and many ghost peaks appear in the spectrum. In Fig. \ref{fig:BER_S2}b and Fig. \ref{fig:BER_S2}c, it is hard to identify the targets from the spectrum, indicating the detection probability of the targets is low in such situation. The CS-ANL1 algorithm, however, is not affected by the demodulation errors and is able to identify the targets correctly.

It is shown via Fig. \ref{fig:BER_S1} and Fig. \ref{fig:BER_S2} that the proposed CS-ANL1 algorithm is robust to demodulation errors, and will have better target detection performance. We also compare the accuracy of the algorithms via a simulation. Note that the algorithms return a bunch of $\tau_k$'s and $f_k$'s, which can be either targets, clutters or false alarms, and a target cannot be identified if there is no $(\tau_k,f_k)$ pair close to its position. To avoid the interference of missed detected targets, we only calculate the range and velocity root-mean-squared error (RMSE) conditioned on correctly identified targets. In our simulation, a target is declared to be correctly identified if there is a detected path whose range error is smaller than $\frac{c}{4 N \Delta f}$ and the velocity error is smaller than $\frac{c}{4 M \bar T f_c}$, which are the grid sizes of the simulated CS-L1 algorithm. For the convenience of our comparison, the power of the three targets are all set as -40dB per sample in this simulation. The powers of the direct path and clutters at each sample are both set as -10dB. To improve the efficiency, we reduce the number of subcarrier frequencies to $N = 16$, and the other settings are the same as those given in Section \ref{sec:setup}. The simulations are run over 100 random realizations in each case.

\begin{figure*}
	\centering
	\subfloat[][]{\includegraphics[width=3in]{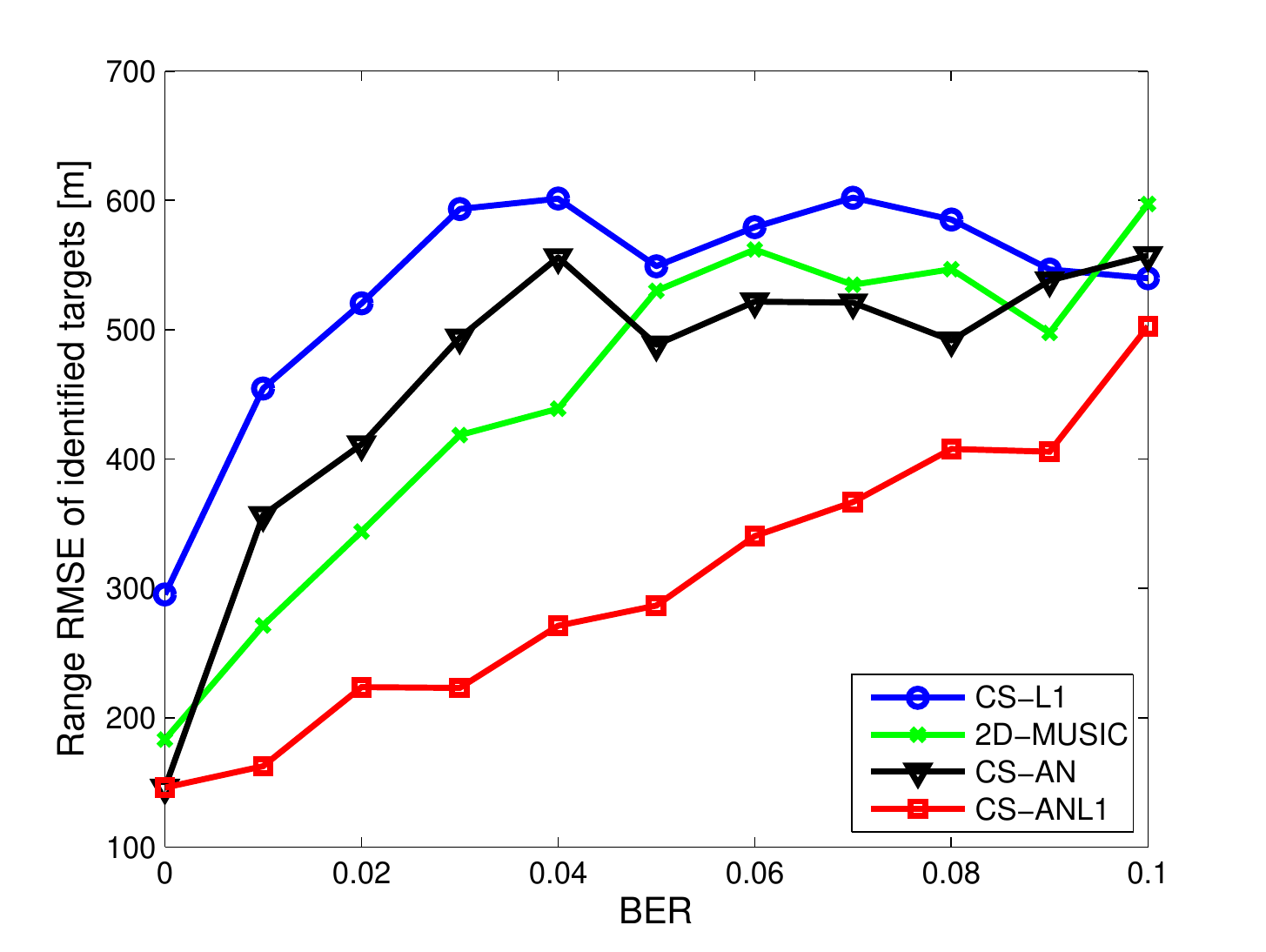}}
	\subfloat[][]{\includegraphics[width=3in]{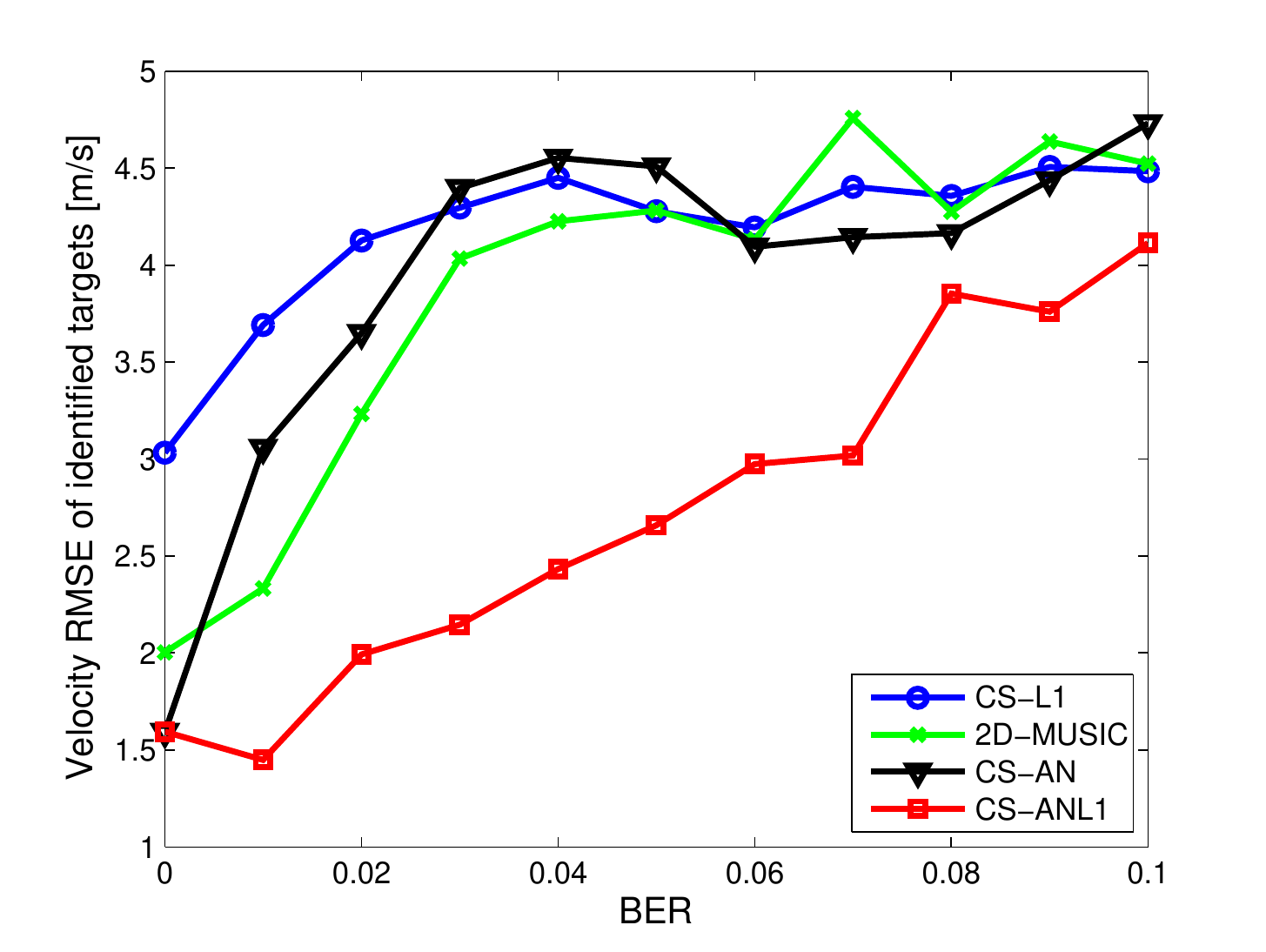}}
	
	\subfloat[][]{\includegraphics[width=3in]{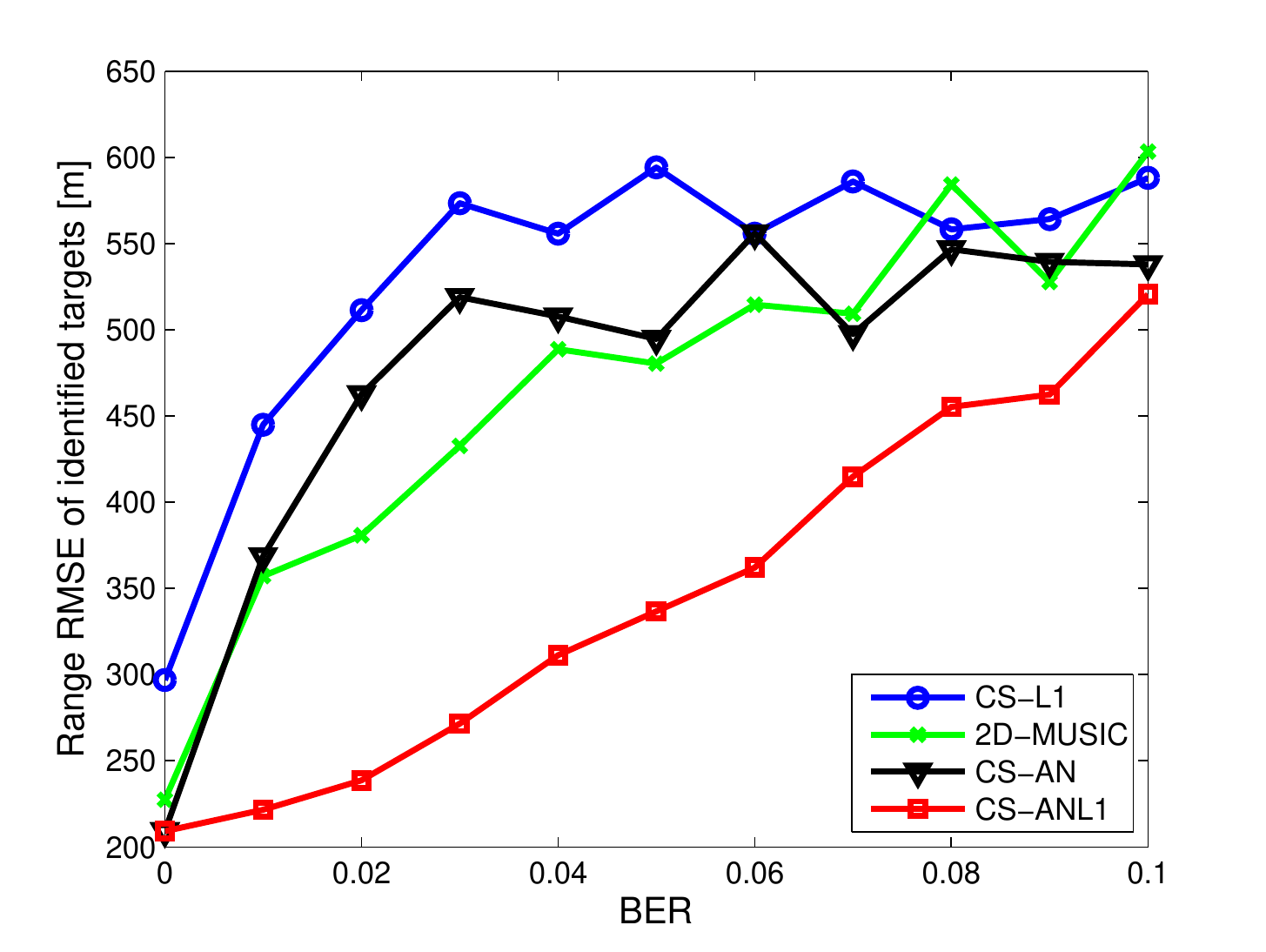}}
	\subfloat[][]{\includegraphics[width=3in]{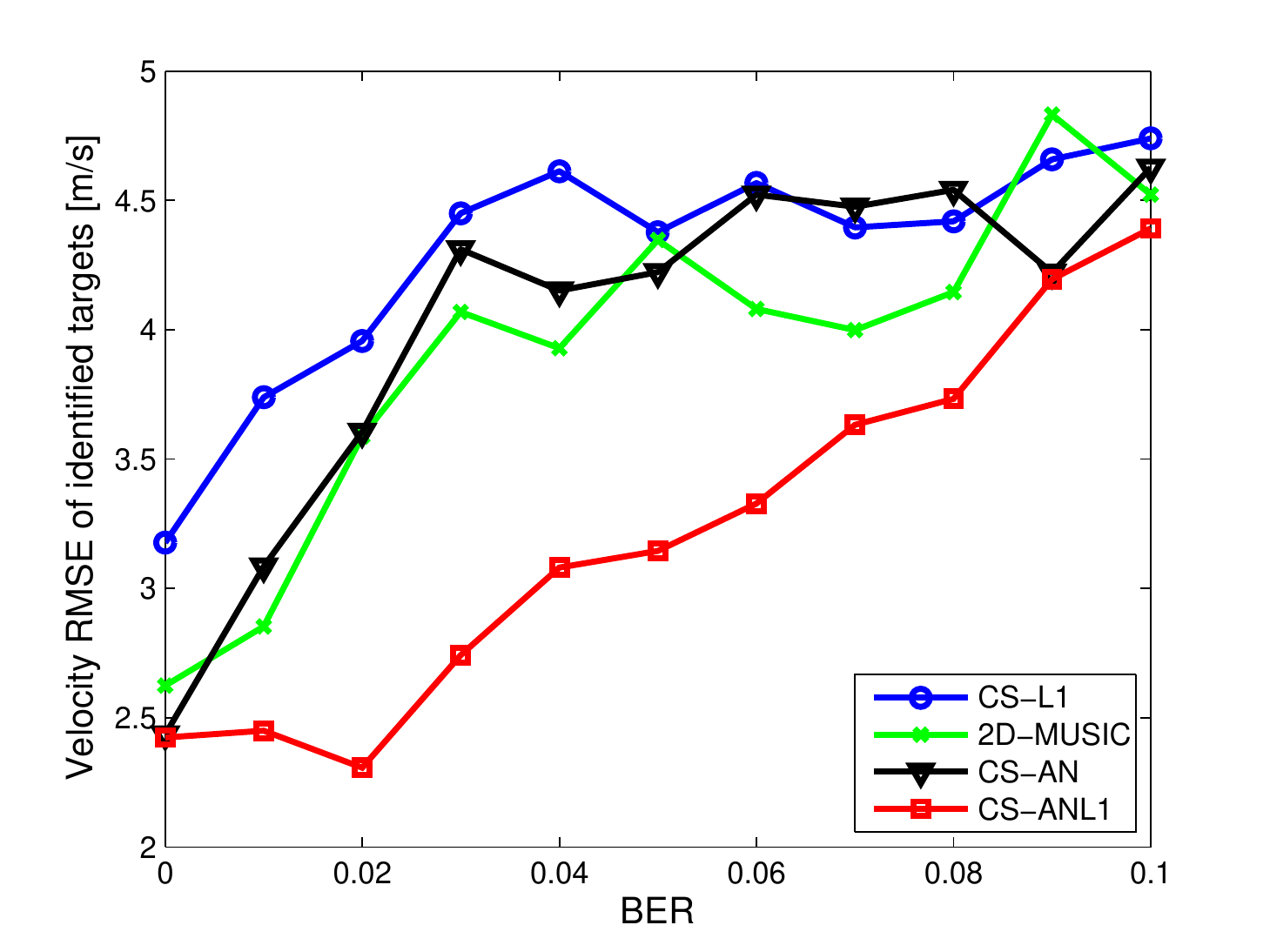}}
	
	\caption{Comparison of the range and velocity accuracy in (a)(b) Scenario 1, and (c)(d) Scenario 2.}
	\label{fig:rmse}
\end{figure*}

 In Fig. \ref{fig:rmse}, we plot the range and velocity RMSE against different BERs. The BER varies from 0 to 0.1 with a step size of 0.01. When BER=0, the accuracies of CS-AN and CS-ANL1 are similar, which are better than those of CS-L1 and 2D-MUSIC. When BER$>$0, the performance of CS-L1, 2D-MUSIC and CS-AN suffers from the impulsive noise and the accuracy decreases as the BER increases. The CS-ANL1 algorithm, however, is able to handle the impulsive noise and provides better accuracy than the other algorithms. It is also worth noting that $\bm{\bar e}$ becomes less sparse as the BER increases. As a result, the performance of the CS-ANL1 algorithm also shows some degration as the BER increases. Specifically, when the BER$=0.1$, the accuracy of CS-ANL1 is close to that of the other algorithms. However, in practical OFDM communications, the demodulation error is usually lower than 0.1.

\section{Conclusions}
In this paper, we have proposed a new high-resolution algorithm for passive radars that make use of OFDM communication signals. It is a compressed sensing (CS)-based algorithm using both the atomic norm and the $\ell_1$-norm to exploit the sparsity of the multipath signal components and the sparsity of the demodulation error, respectively. We have derived an ADMM-based fast algorithm to compute the solution to the formulated CS problem. Simulation results show that the proposed algorithm provides better performance compared to the existing high-resolution methods based on compressed sensing and 2D-MUSIC algorithms, especially in the presence of demodulation errors.  

\section*{Acknowledgements}

We would like to thank the anonymous reviewers and the Associate Editor for their valuable comments and suggestions, which have greatly improved the quality of this paper. We would also like to thank Dr. Peter Willett of University of Connecticut for helpful discussions. This work was supported in part by the U.S. National Science Foundation (NSF) under grant CIF1064575, and in part by the U.S. Office of Naval Research (ONR) under grant N000141410667. 

\appendix
\subsection{Proof of Lemma 2}
Suppose the function $\zeta(\bm{\bar z}, \bm{\bar e}) = \frac{1}{2}{\left\| {{\bm{\bar r}} - \bm{\bar e} - \bm{\tilde S}\bm{\bar z}} \right\|_2^{2}} + \lambda \| \bm{\bar z} \|_{\cal A} + \mu \| \bm{\bar e} \|_1$ is minimized at $(\bm{\hat z},\bm{\hat e})$. 
The following lemma states the optimality condition: 
\begin{lemma}
	$(\bm{\hat z},\bm{\hat e})$ is the solution of \eqref{eq:atomicnorm} if and only if 
	 \begin{eqnarray}
	 \label{eq:zA}
	 &&\lambda \|\bm{\hat z}\|_{\cal A} = \langle \bm{\bar r} - \bm{\hat e} -\bm{\tilde S}\bm{\hat z}, \bm{\tilde S}\bm{\hat z} \rangle, \\
	 \label{eq:e1}
	 &&\mu \| \bm{\hat e}\|_1 = \langle \bm{\bar r} - \bm{\hat e} -\bm{\tilde S}\bm{\hat z}, \bm{\hat e} \rangle, \\
	 \label{eq:dual1}
	 &&\| \bm{\tilde S}^H \bm{\bar r} - \bm{\tilde S}^H \bm{\hat e} - \bm{\tilde S}^H \bm{\tilde S}\bm{\hat z} \|_{\cal A}^* \leq \lambda, \\
	 \label{eq:dual2}
	 &&\| \bm{\bar r} - \bm{\hat e} - \bm{\tilde S}\bm{\hat z} \|_\infty \leq \mu.
	 \end{eqnarray}
\end{lemma}
The proof of the lemma is given in Appendix B.

Plugging \eqref{eq:zA} and \eqref{eq:e1} into $\zeta(\bm{\hat z},\bm{\hat e})$, we have
\begin{eqnarray}
\zeta(\bm{\hat z}, \bm{\hat e}) &=& \frac{1}{2} \| \bm{\bar r} - \bm{\hat e} -\bm{\tilde S}\bm{\hat z} \|_2^2 + \langle \bm{\bar r} - \bm{\hat e} -\bm{\tilde S}\bm{\hat z}, \bm{\tilde S}\bm{\hat z} \rangle_\mathbb{R} + \langle \bm{\bar r} - \bm{\hat e} -\bm{\tilde S}\bm{\hat z}, \bm{\hat e} \rangle_\mathbb{R}
\end{eqnarray}
Note that the duality gap vanishes at $(\bm{\hat z},\bm{\hat e})$, we have $\zeta(\bm{\hat z}, \bm{\hat e}) = \eta(\bm{\hat \nu})$ where $\eta(\bm{\nu}) \triangleq {\left\langle {{(\bm{\tilde S}^H)^{-1} \bm \nu},\bm{\bar r}} \right\rangle _{\mathbb R}} - \frac{1}{2}\left\| {(\bm{S}^H)^{-1} \bm \nu} \right\|_2^2$ is the objective function of the dual problem. This condition is satisfied if the following condition holds:
\begin{eqnarray}
\label{eq:c1}
	\bm{\hat \nu} = \bm{\tilde S}^H (\bm{\bar r} - \bm{\tilde S} \bm{\hat z} - \bm{\hat e}).
\end{eqnarray} 

Using \eqref{eq:c1}, we can build the connection between $\bm{\hat \nu}$ and $\bm{\hat \Upsilon}$. The Lagrangian of \eqref{eq:AL} is given by
\begin{eqnarray}
\label{eq:L}
{\cal L}(t,\bm U,\bm{\bar z},\bm{\bar e},\bm{\Upsilon},\bm \Theta) &=&\frac{\lambda}{2} \left( u_0(0) + t \right) +\mu \|\bm{\bar e}\|_1 + \frac{1}{2} \|\bm{\bar r}-\bm{\bar e} - \bm{\tilde S} \bm{\bar z} \|_2^2 \nonumber \\
&& + \left\langle {{\bm \Upsilon},\bm \Theta  - \left[ {\begin{array}{*{20}{c}}
		{{\cal T}(\bm U)}& \bm{\bar z}\\
		\bm{\bar z}^H & t
		\end{array}} \right]} \right\rangle + \mathbb{I}_\infty(\bm \Theta \succeq 0).
\end{eqnarray}
The Karush-Kuhn-Tucker (KKT) conditions \cite{boyd2004convex} states that the gradient of \eqref{eq:L} with respect to $\bm{\bar z}$ vanishes at $\bm{\bar z} = \bm{\hat z}$. Letting $\nabla_{\bm{\bar z}} {\cal L} = 0$, we have
\begin{eqnarray}
\label{eq:c2}
\bm{\tilde S}^H(\bm{\tilde S}\bm{\bar z} + \bm{\hat e} - \bm{\hat r}) - 2 \bm{\hat \upsilon}_1 = 0.
\end{eqnarray}
Combining \eqref{eq:c1} and \eqref{eq:c2} leads to the relation of $\bm{\hat \nu} = -\frac{\bm{\hat \upsilon_1}}{2}$.

\subsection{Proof of Lemma 3}
As $\zeta(\bm{\hat z},\bm{\hat e})$ is minimized at $(\bm{\hat z},\bm{\hat e})$, for all $\beta \in (0,1)$ and all $(\bm{\bar z}, \bm{\bar e})$, we have
 \begin{eqnarray}
 \zeta(\bm{\hat z}+\beta (\bm{\bar z} - \bm{\hat z}), \bm{\hat e} + \beta(\bm{\bar e} - \bm{\hat e})) \geq \zeta(\bm{\hat z}, \bm{\hat e}),
 \end{eqnarray}
 or equivalently,
 \begin{eqnarray}
 && \beta^{-1} \lambda (\| \bm{\hat z} + \beta (\bm{\bar z} - \bm{\hat z}) \|_{\cal A} - \|\bm{\hat z} \|_{\cal A}) + \beta^{-1} \mu (\| \bm{\hat e} + \beta (\bm{\bar e} - \bm{\hat e}) \|_1 - \|\bm{\hat e} \|_1) \nonumber \\
 &&\geq \langle \bm{\bar r} - \bm{\hat e} -\bm{\tilde S}\bm{\hat z}, \bm{\tilde S}\bm{\bar z} - \bm{\tilde S}\bm{\hat z} \rangle_\mathbb{R} + \langle \bm{\bar r} - \bm{\hat e} -\bm{\tilde S}\bm{\hat z}, \bm{\bar e} - \bm{\hat e} \rangle_\mathbb{R} - \frac{1}{2} \beta \|\bm{\bar e} -\bm{\hat e} + \bm{\tilde S} \bm{\bar z} - \bm{\tilde S} \bm{\hat z} \|_2^2.
 \end{eqnarray}
Since $\| \cdot \|_{\cal A}$ and $\| \cdot \|_1$ are convex, we have
\begin{eqnarray}
\label{eq:neq0}
\lambda( \|\bm{\bar z}\|_{\cal A} - \|\bm{\hat z}\|_{\cal A}) + \mu( \| \bm{\bar e} \|_1 - \| \bm{\hat e}\|_1 ) \geq \beta^{-1} \lambda (\| \bm{\hat z} + \beta (\bm{\bar z} - \bm{\hat z}) \|_{\cal A} - \|\bm{\hat z} \|_{\cal A}) + \beta^{-1} \mu (\| \bm{\hat e} + \beta (\bm{\bar e} - \bm{\hat e}) \|_1 - \|\bm{\hat e} \|_1) 
\end{eqnarray}
Letting $\beta \to 0$, we notice that $(\bm{\hat z},\bm{\hat e})$ minimizes $\zeta(\bm{\bar z},\bm{\bar e})$ if
\begin{eqnarray}
\label{eq:neq}
\lambda( \|\bm{\bar z}\|_{\cal A} - \|\bm{\hat z}\|_{\cal A}) + \mu( \| \bm{\bar e} \|_1 - \| \bm{\hat e}\|_1 ) \geq \langle \bm{\bar r} - \bm{\hat e} -\bm{\tilde S}\bm{\hat z}, \bm{\tilde S}\bm{\bar z} - \bm{\tilde S}\bm{\hat z} \rangle_\mathbb{R} + \langle \bm{\bar r} - \bm{\hat e} -\bm{\tilde S}\bm{\hat z}, \bm{\bar e} - \bm{\hat e} \rangle_\mathbb{R} .
\end{eqnarray}
 
However if \eqref{eq:neq} holds, then for all $\bm{\bar z}$ and $\bm{\bar e}$
\begin{eqnarray}
&&\frac{1}{2} \left\| \bm{\bar r} - \bm{\bar e} - \bm{\tilde S} \bm{\bar z} \right\|_2^2 + \lambda \| \bm{\bar z} \|_{\cal A} + \mu \| \bm{\bar e} \|_1 \nonumber \\
&&\geq \frac{1}{2} \left\| \bm{\bar r} - \bm{\hat e} - \bm{\tilde S} \bm{\hat z} + \bm{\hat e} - \bm{\bar e} + \bm{\tilde S}\bm{\hat z} - \bm{\tilde S} \bm{\bar z} \right\|_2^2  + \langle \bm{\bar r} - \bm{\hat e} -\bm{\tilde S}\bm{\hat z}, \bm{\tilde S}\bm{\bar z} - \bm{\tilde S}\bm{\hat z} \rangle_\mathbb{R} \nonumber \\
&& ~~ + \langle \bm{\bar r} - \bm{\hat e} -\bm{\tilde S}\bm{\hat z}, \bm{\bar e} - \bm{\hat e} \rangle_\mathbb{R} + \lambda\|\bm{\hat z}\|_{\cal A} + \mu\|\bm{\hat e}\|_1 \nonumber \\
&& = \frac{1}{2} \left\| \bm{\bar r} - \bm{\hat e} - \bm{\tilde S} \bm{\hat z} \right\|_2^2 + \lambda\|\bm{\hat z}\|_{\cal A} + \mu\|\bm{\hat e}\|_1
\end{eqnarray}
implying $\zeta(\bm{\bar z}, \bm{\bar e}) \geq \zeta(\bm{\hat z}, \bm{\hat e})$. Hence, \eqref{eq:neq} is necessary and sufficient for the optimum solution. We then rewrite \eqref{eq:neq} as
 \begin{eqnarray}
 \label{eq:neq2}
 && \inf_{\bm{\bar z}} \left(\lambda \|\bm{\bar z}\|_{\cal A} - \langle \bm{\bar r} - \bm{\hat e} -\bm{\tilde S}\bm{\hat z}, \bm{\tilde S}\bm{\bar z} \rangle_\mathbb{R} \right) + \inf_{\bm{\bar e}} \left( \mu \| \bm{\bar e} \|_1   - \langle \bm{\bar r} - \bm{\hat e} -\bm{\tilde S}\bm{\hat z}, \bm{\bar e} \rangle_\mathbb{R} \right)  \nonumber \\
 &&\geq \lambda \|\bm{\hat z}\|_{\cal A} + \mu \| \bm{\hat e}\|_1 - \langle \bm{\bar r} - \bm{\hat e} -\bm{\tilde S}\bm{\hat z}, \bm{\tilde S}\bm{\hat z} \rangle_\mathbb{R} - \langle \bm{\bar r} - \bm{\hat e} -\bm{\tilde S}\bm{\hat z}, \bm{\hat e} \rangle_\mathbb{R}.
 \end{eqnarray}
 By the definition of the dual atomic norm, we have
 \begin{eqnarray}
\inf_{\bm{\bar z}} \lambda \|\bm{\bar z}\|_{\cal A} - \langle \bm{\bar r} - \bm{\hat e} -\bm{\tilde S}\bm{\hat z}, \bm{\tilde S}\bm{\bar z} \rangle &=& \left\{ \begin{array}{l}
 0, \| \bm{\tilde S}^H \bm{\bar r} - \bm{\tilde S}^H \bm{\hat e} - \bm{\tilde S}^H \bm{\tilde S}\bm{\hat z} \|_{\cal A}^* \leq \lambda, \\
 -\infty, \| \bm{\tilde S}^H \bm{\bar r} - \bm{\tilde S}^H \bm{\hat e} - \bm{\tilde S}^H \bm{\tilde S}\bm{\hat z} \|_{\cal A}^* > \lambda,
 \end{array} \right. \\
\inf_{\bm{\bar e}} \mu \|\bm{\bar e}\|_1 - \langle \bm{\bar r} - \bm{\hat e} -\bm{\tilde S}\bm{\hat z}, \bm{\bar e} \rangle &=& \left\{ \begin{array}{l}
 0, \| \bm{\bar r} - \bm{\hat e} - \bm{\tilde S}\bm{\hat z} \|_\infty \leq \mu, \\
 -\infty, \| \bm{\bar r} - \bm{\hat e} - \bm{\tilde S}\bm{\hat z} \|_\infty > \mu.
 \end{array} \right. 
 \end{eqnarray}
If $\| \bm{\tilde S}^H \bm{\bar r} - \bm{\tilde S}^H \bm{\hat e} - \bm{\tilde S}^H \bm{\tilde S}\bm{\hat z} \|_{\cal A}^* > \lambda$ or $\| \bm{\bar r} - \bm{\hat e} - \bm{\tilde S}\bm{\hat z} \|_\infty > \mu$, the left side of \eqref{eq:neq2} is going to $-\infty$. Hence, $(\bm{\hat z},\bm{\hat e})$ have to satisfy the constraints $\| \bm{\tilde S}^H \bm{\bar r} - \bm{\tilde S}^H \bm{\hat e} - \bm{\tilde S}^H \bm{\tilde S}\bm{\hat z} \|_{\cal A}^* \leq \lambda$ and $\| \bm{\bar r} - \bm{\hat e} - \bm{\tilde S}\bm{\hat z} \|_\infty \leq \mu$, which indicates
\begin{eqnarray}
\label{eq:condition1}
&&\lambda \|\bm{\hat z}\|_{\cal A} \geq \langle \bm{\bar r} - \bm{\hat e} -\bm{\tilde S}\bm{\hat z}, \bm{\tilde S}\bm{\hat z} \rangle_\mathbb{R}, \\
\label{eq:condition2}
&& \mu \| \bm{\hat e}\|_1 \geq \langle \bm{\bar r} - \bm{\hat e} -\bm{\tilde S}\bm{\hat z}, \bm{\hat e} \rangle_\mathbb{R}.
\end{eqnarray}
Plugging $\inf_{\bm{\bar z}} \left(\lambda \|\bm{\bar z}\|_{\cal A} - \langle \bm{\bar r} - \bm{\hat e} -\bm{\tilde S}\bm{\hat z}, \bm{\tilde S}\bm{\bar z} \rangle_\mathbb{R} \right) = 0$ and $ \inf_{\bm{\bar e}} \left( \mu \| \bm{\bar e} \|_1   - \langle \bm{\bar r} - \bm{\hat e} -\bm{\tilde S}\bm{\hat z}, \bm{\bar e} \rangle_\mathbb{R} \right) = 0$ into \eqref{eq:neq2}, we have 
\begin{eqnarray}
\label{eq:condition3}
\lambda \|\bm{\hat z}\|_{\cal A} + \mu \| \bm{\hat e}\|_1 \leq \langle \bm{\bar r} - \bm{\hat e} -\bm{\tilde S}\bm{\hat z},  \bm{\tilde S}\bm{\hat z} \rangle_\mathbb{R} + \langle \bm{\bar r} - \bm{\hat e} -\bm{\tilde S}\bm{\hat z}, \bm{\hat e} \rangle_\mathbb{R}.
\end{eqnarray}
Combining \eqref{eq:condition1}, \eqref{eq:condition2} and \eqref{eq:condition3} results in the relations in \eqref{eq:zA}, \eqref{eq:e1}, \eqref{eq:dual1} and \eqref{eq:dual2}. 

\bibliographystyle{IEEEtran}
\bibliography{database}

\end{document}